\documentclass[fleqn,usenatbib,useAMS]{mnras}

\usepackage{graphicx}	
\usepackage{amsmath}	
\usepackage{multicol}        
\usepackage{bm}		
\usepackage{pdflscape}	

\usepackage[T1]{fontenc}
\usepackage{ae,aecompl}

\usepackage{newtxtext,newtxmath}

\title[Project Hephaistos: JWST]{Project Hephaistos -- IV. James Webb Space Telescope Observations of Two Dyson Sphere Candidates}

\author[E. Zackrisson et al.]{Erik Zackrisson$^{1}$
\thanks{Contact e-mail: \href{mailto:erik.zackrisson@physics.uu.se}{erik.zackrisson@physics.uu.se}}, Arjan Bik$^{2}$, Anita Ali Asgar$^{1}$, Olivia Curtis$^{3,4,5}$, Jason T Wright$^{3,4,5}$,  Tongtian Ren$^{6}$, \newauthor Roberto J. Assef$^{7}$,  Andrew Blain$^{8}$, Alexis Brandeker$^{2}$, Michael A. Garrett$^{6,9,10}$, Uma Gorti$^{11,12}$, \newauthor Andreas J. Korn$^{1}$,  Priyatam K. Mahto$^{13}$, Armin Nabizadeh$^{1}$
\\
$^{1}$Observational Astrophysics, Department of Physics and Astronomy, Uppsala University, Box 516, SE-751 20 Uppsala, Sweden\\
$^{2}$  Department of Astronomy, Stockholm University, AlbaNova University Centre, 106 91 Stockholm, Sweden\\
$^{3}$ Department of Astronomy \& Astrophysics, 525 Davey Laboratory, The Pennsylvania State University, University Park, PA, 16802, USA\\
$^{4}$ Center for Exoplanets and Habitable Worlds, 525 Davey Laboratory, The Pennsylvania State University, University Park, PA, 16802, USA\\
$^{5}$ Penn State Extraterrestrial Intelligence Center, 525 Davey Laboratory, The Pennsylvania State University, University Park, PA, 16802, USA\\
$^{6}$ Jodrell Bank Centre for Astrophysics, Department of Physics and Astronomy, University of Manchester, Oxford Road, Manchester M13 9PL, UK\\
$^{7}$ Instituto de Estudios Astrof\'isicos, Facultad de Ingenier\'ia y Ciencias, Universidad Diego Portales, Av. Ej\'ercito Libertador 441, Santiago, Chile\\
$^{8}$ School of Physics and Astronomy, University of Leicester, University Road, Leicester LE1 7RH, UK \\
$^{9}$ Leiden Observatory, Leiden University, PO Box 9513, NL-2300 RA Leiden, the Netherlands\\
$^{10}$ University of Malta, Institute of Space Sciences and Astronomy, Msida, MSD2080, Malta\\
$^{11}$ NASA Ames Research Center, Moffett Field, CA 94035, USA\\
$^{12}$ Carl Sagan Center, SETI Institute, Mountain View, CA 94043, USA\\
$^{13}$ Departamento de Astronom\'ia, Universidad de Chile, Camino El Observatorio 1515, Las Condes, Santiago, Chile}

\date{Last updated XXX; in original form XXX}

\pubyear{2026}

\begin{document}

\label{firstpage}
\pagerange{\pageref{firstpage}--\pageref{lastpage}}
\maketitle

\begin{abstract}
We report on JWST/MIRI imaging and spectroscopy of two M-dwarf stars previously singled out by project Hephaistos as potential Dyson-sphere candidates (their candidates D and E) due to the presence of excess flux at mid-infrared wavelengths. We find that the infrared excess does not originate from Dysonian megastructures, or other radiation mechanisms close to these stars, but from background galaxies projected within $\sim 1$ arcsec of the M dwarfs, thereby confusing previous mid-infrared photometry obtained with the WISE telescope. The candidate D background galaxy lies at redshift $z\approx 0.9$, appears point-source dominated in imaging and has a mid-infrared spectrum consistent with being a Hot Dust Obscured Galaxy (Hot DOG). The candidate E background galaxy lies at $z\approx 0.4$, displays an extended morphology with bright knots and a spectrum consistent with a dusty starburst.   
\end{abstract}





\section{Introduction}
\label{sec:intro}
Dyson spheres \citep{Dyson60} are hypothetical, artificial megastructures that technologically advanced civilizations may build to harvest radiation energy from stars \citep[for a review, see][]{Wright20}, potentially for the purpose of powering vessels, supercomputers or habitats, or to send messages over interstellar distances. The primary observable signature of Dyson spheres is the ``waste heat'' emitted from these structures, which for a wide range of Dyson sphere temperatures would produce excess flux in the infrared (IR) wavelength range. Other potential signatures, more dependent on the size of the absorbing elements of the sphere and the fraction of the star light absorbed, include dimming of direct star light and variability due to the transit of the absorbing elements across the stellar disk. In Project Hephaistos paper I, \citet{Suazo22} studied the infrared excess of stars with Gaia and WISE data to place conservative upper limits on the prevalence of partial Dyson spheres at distances out to 5 kpc. These limits were, however, relatively weak  -- except in the case of Dyson spheres with very high covering fractions -- since there are several astrophysical processes that can naturally produce excess IR flux, with dust-enshrouded young stars found to be one of the most severe contaminants. In Project Hephaistos paper II, \citet{Suazo24} used a sample of $\approx 5\times10^{6}$ stars with Gaia, 2MASS and WISE data to
 single out a small number of potential Dyson sphere candidates, all of them M dwarfs, which exhibit significant IR excess in the WISE W3 (12 $\mu$m) and W4 bands (22 $\mu$m), while at the same time not showing any obvious signs of stellar youth. \citet{Suazo24b} and Korn et al. (2026, in prep.) also present optical medium-resolution spectra for some of these, which corroborate the conclusion that these stars -- at optical wavelengths -- appear to be perfectly normal main-sequence M dwarfs, without strong H$\alpha$ emission (which would be a signature of gas accretion, i.e. youth). 
 
Main-sequence stars may still exhibit infrared excess flux due to the presence of a debris disk, either left over from the formation of protoplanets, or as a result of late-occurring collisions within a planetary system \citep[for a review, see][]{Hughes18}. The \citet{Suazo24} Dyson-sphere candidates have inferred IR-to-bolometric luminosity ratios $L_\mathrm{IR}/L_\mathrm{tot}$ in the $\approx 0.07-0.17$ range, which could be consistent with extreme debris disks ($L_\mathrm{IR}/L_\mathrm{tot} \gtrsim 0.01$), believed to be caused by signatures of giant impacts during the final assembly of rocky planets, the outcome of a late dynamical instability of a planetary system, or planet-moon collisions due to tidal evolution. While many extreme debris disks have been detected around FGK stars \citep[e.g.][]{Moor21,Moor24}, the only extreme debris disks detected around M dwarfs are the ``Peter Pan disks'' (debris disks that ``refuse to grow up'') presented by \citet{Silverberg20}. These objects do, however, exhibit H$\alpha$ emission, so if the infrared excess seen in the  \citet{Suazo24} sample is caused by extreme debris disks, they would need to belong to a class not previously seen around M dwarfs.

An alternative explanation for the infrared excess of the \citet{Suazo24} objects could be that the infrared radiation is not produced anywhere near these objects, but comes from extremely red, infrared-bright background objects, which by chance are projected very close to these stars, and contribute negligible flux at optical and near-IR wavelengths  \citep{Blain24,Ren24,Ren25}. Given the poor angular resolution of the WISE W3 and W4 imaging data
(point-spread function full-width at half-maximum, FWHM, $\approx 6$ arcsec in W3 and $\approx 12$ arcsec in W4; \citealt{Lang14}), this may not be immediately obvious from a visual inspection of the data. If the signal-to-noise ratios of the infrared observations are sufficient, it may however be possible to detect this as a shift in the centroid of the stellar image when going from the shortest-wavelength WISE bands (where the star is expected to dominate) to the longest-wavelength bands (where the background object would dominate). \citet{Suazo24} did attempt to constrain the centroid shifts, arguing that the WISE centroid shifts were consistent with a single source within the error bars. However, a subsequent analysis of centroid shifts from the optical-to-mid-IR \citep{Ren26} did provide evidence for centroid shifts in the majority of the \citet{Suazo24} candidates. Additional support for the background-interloper hypothesis come from the detection of radio emission close to several of the \citet{Suazo24} candidates \citep{Ren24,Ren25}. Since intrinsic, persistent radio emission from M dwarfs at detectable levels is rare \citep[$\approx 0.5\%$ in the sample of][]{Callingham21}, this strongly suggests that there is a background active galactic nucleus close to the position of at least some of these candidates. \citet{Ren24,Ren25} and \citet{Blain24} suggest that these objects may be Hot Dust Obscured Galaxies (Hot DOGs; \citealt{Eisenhardt12}), a rare population of dust-obscured quasars with hot dust \citep[observed temperatures ranging from $\sim70$ K to $\sim 450$ K;][]{Tsai15,Sun24} and very red spectral energy distributions across the WISE bands \citep{Wu12}. While the Hot DOG population seems to have peaked at redshift $z\approx 2$--3 \citep{Assef15}, a few examples of Hot DOGs at $z<0.5$ have also been identified \citep{Li25}.

In this paper, we present James Webb Space Telescope (JWST) imaging and spectroscopy using the JWST Mid-Infrared Instrument (MIRI) for two of the \citet{Suazo24} Dyson-sphere candidates without radio detections (candidates D and E in their notation) to further constrain the origin and nature of their infrared excess fluxes. Section~\ref{sec:observations} presents the observational data used. In section~\ref{sec:SED_modelling} we discuss how the photometry from the nearby companions was deblended, and in section~\ref{sec:nature of background objects} our conclusions regarding the mid-infrared (5--25 $\mu$m) properties of our targets. Section~\ref{sec:discussion} discusses the impact of our findings on future searches for Dyson spheres and the benefit for non-technosignature astronomy from observations of this type. Section~\ref{sec:summary} summarizes our findings.

\section{Observational data}
\label{sec:observations}
Project Hephaistos candidates D (RA 23:27:51.3277, Dec +05:06:26.49 in international celestial reference system, ICRS, coordinates at epoch 2000) and E (RA 04:02:7.7601, Dec -10:54:41.21) were observed by JWST on July 28 and Sept 9, 2025 as part of JWST cycle 4 programme GO 7199 \citep{Zackrisson25}. MIRI imaging of the candidates a nd the surrounding field in filters F560W, F1000W and F1500W were obtained with 4 dithers and total exposure times of $\approx 1077$ s per filter using the FASTR1 readout mode, in parallel with MIRI medium-resolution spectroscopy (MRS) of background fields in channels 1--4 and  wavelength ranges A (short), B (medium) and C (long), with exposure times $\approx 1147$ s per wavelength range. MIRI/MRS observations of candidates D and E where executed in direct connection to the imaging and background observations. The MRS observations were obtained using the SLOWR1 readout mode with the same settings in on-target and  background MRS observations, except for the dither pattern (4-point dithers were used in both cases, but with dither patterns optimized for a point-source and for background, respectively). 

\subsection{Archival data}
In addition to the JWST data, we make use of archival photometry for candidates D and E in filters $BP$ ($\approx 0.51\ \mu$m), $G$ ($\approx 0.62\ \mu$m) and $RP$ ($\approx 0.78\ \mu$m) from Gaia Data Release 3 \citep{Gaia_mission_paper,Gaia_DR3}; filters $i$ ($\approx 0.75\ \mu$m), $y$ ($\approx 0.87\ \mu$m) and $z$ ($\approx 0.96\ \mu$m) from the Panoramic Survey Telescope and Rapid Response System  (Pan-STARRS) PS1 survey \citep{Chambers16}; filters $J$ ($\approx 1.2\ \mu$m), $H$ ($\approx 1.7\ \mu$m) and $K$ ($\approx 2.2\ \mu$m) filters from the 2MASS \citep{Skrutskie06}  All-Sky Catalog of Point Sources \citep{Vizier_2MASS} and in WISE \citep{Wright10} filters $W1$ ($\approx 3.4\ \mu$m), $W2$ ($\approx 4.6\ \mu$m), $W3$ ($\approx 12\ \mu$m) and $W4$ ($\approx 22\ \mu$m) from the ALLWISE catalog \citep{Vizier_ALLWISE}. We also use $\approx 0.74$--5.0 $\mu$m low-resolution spectra for candidates D and E from SPHEREx \citep{Bock26}, obtained from QR2\footnote{https://doi.org/10.26131/IRSA652} data using the SPHEREx spectrophotometry tool.

\subsection{MIRI imaging and photometry}
\label{subsec:imaging}
The imaging data were processed through the standard JWST calibration pipeline version 1.20.2 \cite{Bushouse2022} using the CRDS context 1464 (software version 13.0.6). We work with the sky-projected mosaics in each filter as science images in units of MJy~sr$^{-1}$, with matched per-pixel uncertainty arrays as our noise model. Since neither field contained a bright, isolated reference star, we elect to generate model point spread functions (PSFs) for each filter with \texttt{stpsf} \citep{Perrin14}. Each model PSF was computed at the detector position of the target over a $135\times135$~pixel field of view with a Gaussian pointing jitter of $\sigma=0.06'$.

\begin{figure*}
\includegraphics[scale=0.4]{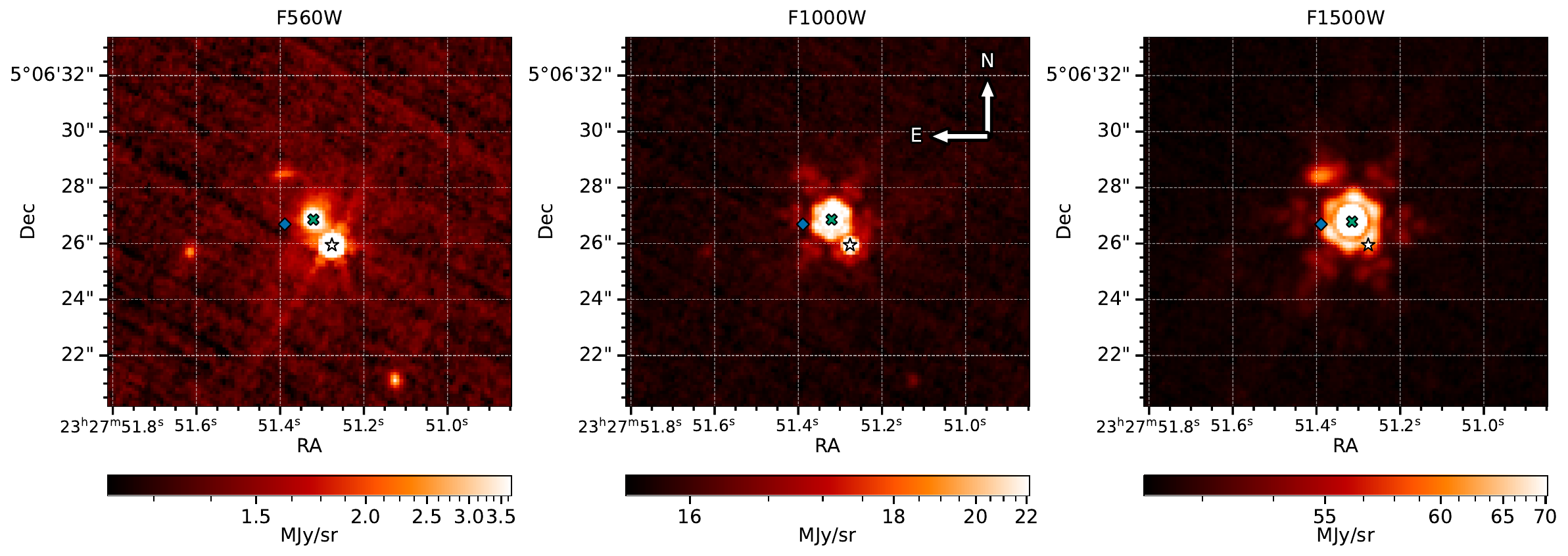}\\
\includegraphics[scale=0.4]{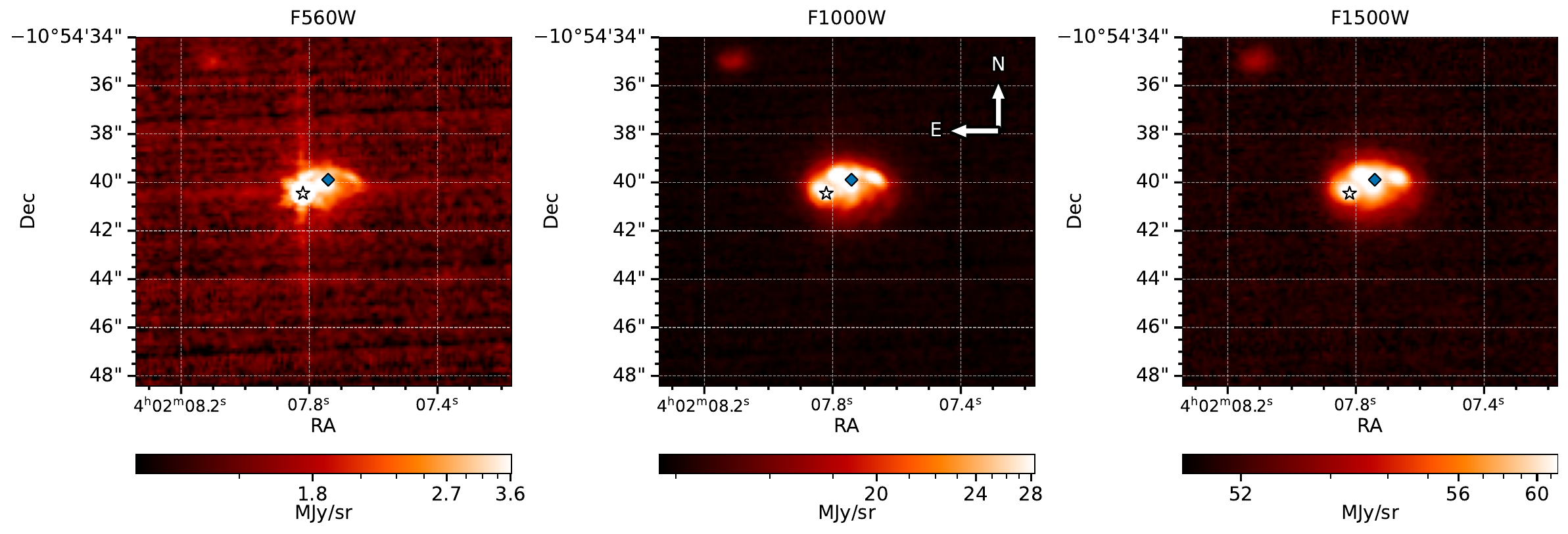}
\caption{Images in the JWST/MIRI F560W, F1000W and F1500W filters (central wavelength $\approx$ 5.6, 10 and 15 $\mu$m) of candidates D (first row) and E (second row).  White stars mark the Gaia positions of the M dwarfs, which in both cases are seen to be located  $\approx 1$ arcsec from the centres of background galaxies, with redshifts $z\approx 0.9$ (candidate D) and $z\approx 0.4$ (candidate E) inferred from the MIRI spectroscopy (Section~\ref{subsec:spectroscopy}). Blue diamonds mark the WISE W3 centroids \citep[$\approx 12\ \mu$m;][]{Ren26}, which are displaced from the Gaia positions of the M dwarfs toward the background galaxies (by $0.75\pm 0.22$ arcsec for candidate D and $1.50\pm 0.24$ arcsec for candidate E), as expected if the galaxies dominate the flux in this band. This illustrates the source confusion responsible for the apparent mid-IR excess of these stars. Green crosses mark the centroids of the candidate D background galaxy derived from our PSF fitting. We do not plot a galaxy centroid for candidate E as the severe blending prevents a reliable fit. The arrows on the compass rose are $2''$ in length. All images have been calibrated to Gaia  centroids that have been shifted to J2025 coordinates using the proper motions of the stars. The background galaxies are subdominant in F560W but come to dominate the light in F1000W and F1500W. In the case of candidate D, the light from the background galaxy is dominated by an unresolved point-source. While there is an extended object to the northeast of the point source (most clearly seen in F560W and F1500W), this appears to be an interloper at a different redshift (Section~\ref{subsec:features and z}). In the case of candidate E, the background galaxy appears extended in all images, and displays several bright clumps in F1000W and F1500W. One of these regions  happens to overlap with the position of the M dwarf (Section~\ref{subsec:spectroscopy}) and dominates the light from this position in both filters. Hence, the light from the M dwarf can for candidate E only be clearly distinguished in F560W. From these images, we conclude that the excess mid-IR flux seen in WISE data for Dyson-sphere candidates D and E can be attributed to contamination by mid-IR bright background galaxies that appear blended with the M-dwarf stars when observed at the angular resolution of WISE.} 
\label{fig:MIRI_images}		
\end{figure*}

In Figure~\ref{fig:MIRI_images}, we display MIRI F560W, F1000W and F1500W images of candidates D and E. In both cases, background objects are evident at distances $\approx 1$ arcsec from the M-dwarf stars. In the case of D, the background object appears as unresolved point-source, and in E as an extended object with several bright regions appearing in F1000W and F1500W. Since MIRI probes the Rayleigh-Jeans tail of the M-dwarf spectra, these stars appear progressively fainter at longer wavelengths. For both candidates D and E, the M dwarf dominates the total light in F560W, wheras the background objects dominate the light in F1000W and F1500W. In the case of candidate D, the M dwarf can be clearly seen in F560W and F1000W, but is largely invisible against the background galaxy in F1500W. For candidate E, the M dwarf is visible in F560W, but happens to overlap with one of the bright regions of the background galaxy in F1000W and F1500W (see Section~\ref{subsec:spectroscopy}). From these images alone, it is clear that the mid-IR excess noted by \citet{Suazo24} is dominated by the contamination of background objects, and not caused by Dyson spheres or debris disks, as these would appear unresolved in the MIRI images (See Section~\ref{sec:discussion}) -- hence, no secondary source would be expected to be visible if these were the primary mechanisms behind the mid-IR excess. 

\subsection{MIRI spectroscopy}
\label{subsec:spectroscopy}
The MIRI-MRS spectra were reduced using version V1.18.0 of the JWST pipeline \citep{Bushouse2025} with CRDS context 1364. Both the on-source and background observations were processed identically with Stage 1 of the pipeline using the default settings.  Stage 2 of the pipeline was run using the {\tt clean\_showers} in the straylight step to remove cosmic ray showers not corrected for in the jump step of stage 1. We use the output of stage 2 of the background observations to calculate a master background by median averaging the cal files, while applying a sigma clipping rejecting 3 sigma outliers. This master background was subsequently subtracted from each of the cal files of the on-source observations. We ran stage 3 of the pipeline on the background subtracted cal files in order to create a reduced data cube. 

Inspection of the final data cubes shows that a residual sky background remained present, especially at the longer wavelength. Therefore we applied an additional background subtraction when extracting the integrated spectra. 
Using \texttt{CUBEVIZ} \citep{Jdaviz}, for each object, an elliptical region was selected to encompass the galaxy and star at the wavelength where they showed the largest spatial extent, resulting in major and minor axes of 1.4" $\times$ 1.4" (candidate D) and 2.0" $\times$ 1.6" (candidate E). We then chose the background aperture as a circular annulus region, with an inner radius of 1.6" and 2.2", and an outer radius of 2.5" and 3.1" for candidates D and E respectively, in channel 3. 
A manual inspection proved the need to (1) remove an unknown interloper from the northeast side of candidate D from the background aperture, and (2) center the ellipse around the star, and reduce its size for candidate E, channel 1, in order to provide background apertures without artifacts. The spectra were subsequently extracted. Synthetic photometry based on these spectra are well matched to the observed photometry of Table \ref{table:photometry}. 

\begin{figure*}
\includegraphics[width=\columnwidth]{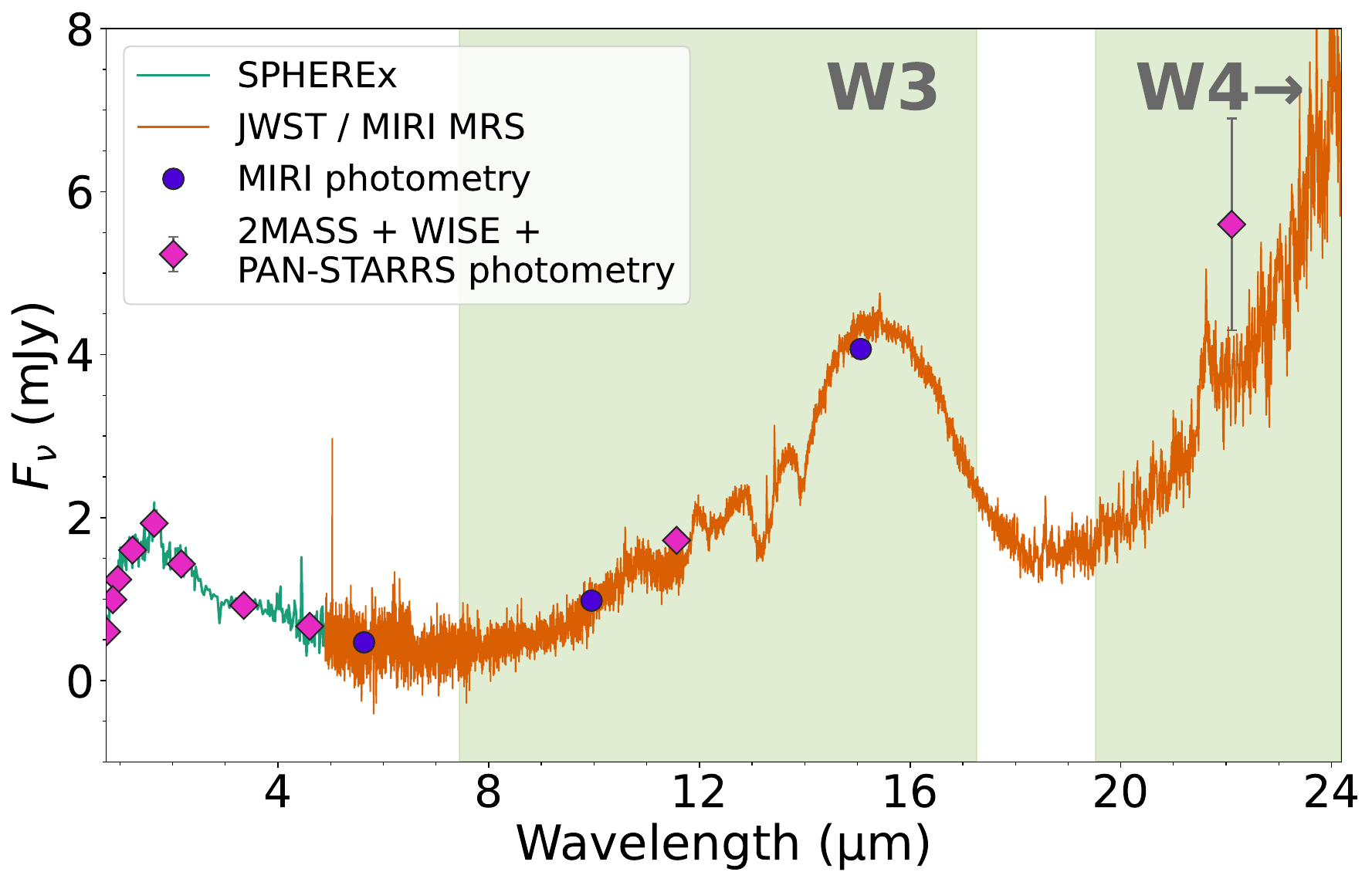}
\includegraphics[width=\columnwidth]{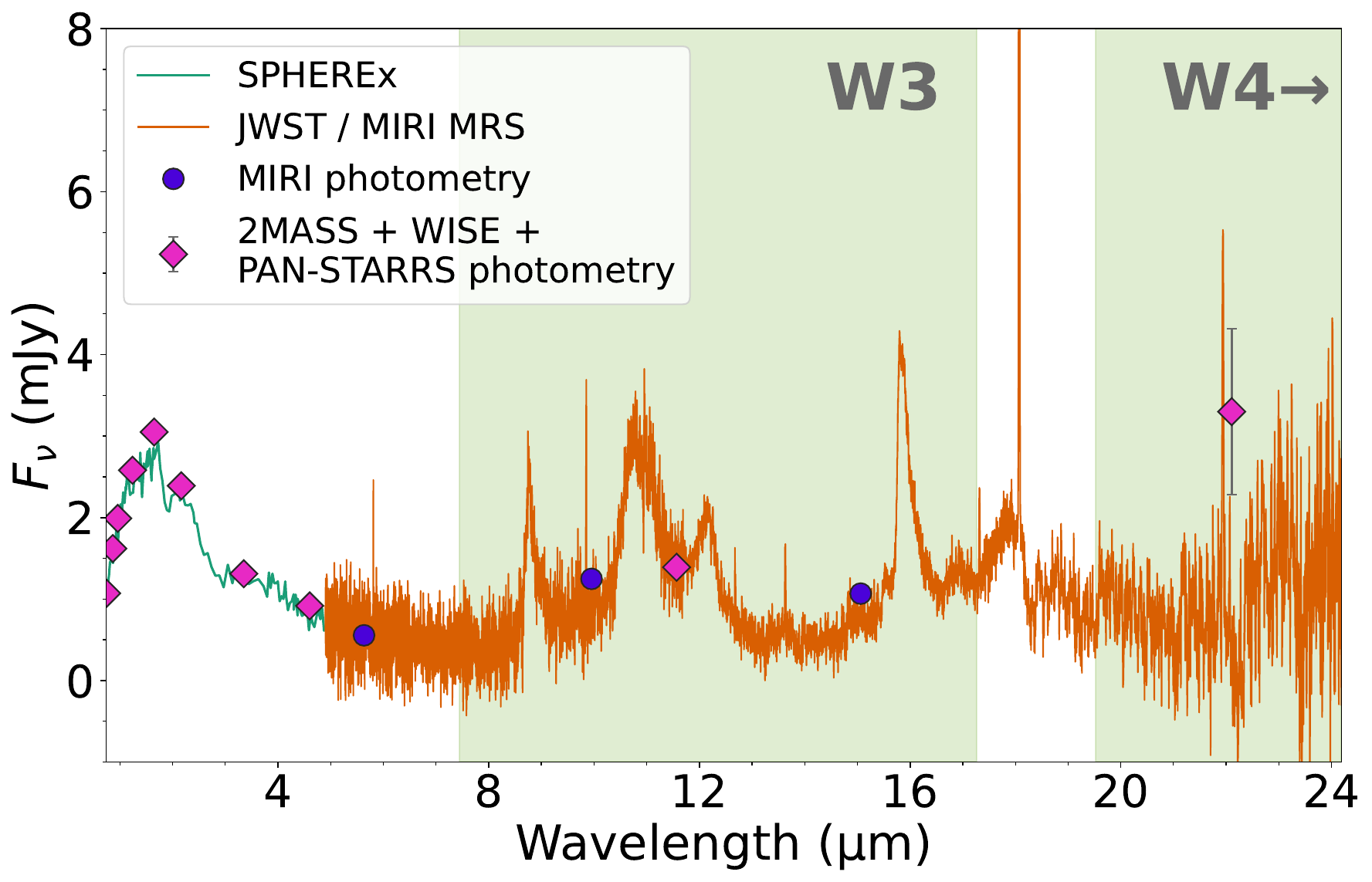}
\caption{Observed spectra and photometry of candidate D (left panel) and E (right panel). SPHEREx spectra at 0.74--5.0 $\mu$m are marked in teal and JWST/MIRI spectra at 4.9--24.0 $\mu$m are marked in orange. Photometric data points from PAN-STARRS, 2MASS and WISE are indicated by pink diamonds. Photometric data points based on MIRI F560W, F1000W and F1500W images are shown as blue circles. Error bars are typically smaller than the symbols, except in the case of the observed WISE $W$4 photometry. The spectrum ends at $24.2 \upmu$m as the high noise levels in sub-channel 4C hide any spectral features.}
\label{fig:MIRI_spectra _obsframe}		
\end{figure*}

In Figure~\ref{fig:MIRI_spectra _obsframe}, we show the observed MIRI spectra from candidates D and E at 4.9--24 $\mu$m, along with the SPHEREx spectra at 0.74--5.0 $\mu$m, the observed total (star + background object) MIRI photometry and archival photometry from PAN-STARRS (filters {\it yzi}), 2MASS ({\it JHK}) and WISE ($W$1, $W$2, $W$3, $W$4). We find continuum levels of the SPHEREx and MIRI spectra agree with each other at the overlap region around $\approx 5$ $\mu$m, and with most of the available photometry, which indicates that the MIRI spectrum is well-calibrated \citep{Law2025}. However, we find an offset between the MIRI MRS spectra and the WISE W3/W4 filters for both candidates, as seen in Table \ref{tab:offsets}, the origin(s) of which have not been made clear with the available data. Most notably, there is a $\sim4 \sigma$ discrepancy in the W4 filter for candidate E. Color-correction effects of the wide W3 filter were investigated by applying the correction factor corresponding to the best-fit power-law function, which increased the offset for candidate D. Moreover, there are no signs of a temporal variability between WISE exposures. Differences in aperture size causing flux loss due to the extended nature of the candidates, and inaccurate background subtraction may contribute to the observed discrepancies, especially in the W4 band. While the cause of these offsets remain unclear, the good agreement between the synthetic photometry and MIRI Imaging data show that the extracted spectra capture a majority of the source emission.

\begin{table*}
\centering
\caption{Offsets between synthetic photometry derived from the MIRI MRS spectra, and WISE photometry in the W3/W4 filters. A 10\% red source correction has been applied for the W4 filter.}
\label{tab:offsets}
\begin{tabular}{ccccccccc}
\hline
Candidate & W3$_S$\textsuperscript{a} & W3$_C$\textsuperscript{b} & 
$\Delta$W3\textsuperscript{c} & $|\Delta \sigma_{W3}|$ \textsuperscript{d}  & W4$_S$\textsuperscript{a} & W4$_C$\textsuperscript{b} & 
$\Delta$W4\textsuperscript{c} &
$|\Delta \sigma_{W4}|$ \textsuperscript{d} \\
\hline
D & $15.55 \pm 0.06$ & $15.82 \pm 0.10$   & $-0.27 \pm 0.12$ & $\sim 2$ 
  & $14.85 \pm 0.06$ & $14.54 \pm 0.23$ & $0.31  \pm 0.24$ & $\sim 1$ \\
E & $16.30 \pm 0.06$ & $16.05 \pm 0.11$   & $0.25  \pm 0.13$ & $\sim 2$         & $16.36 \pm 0.06$ & $15.11 \pm 0.30$   & $1.25  \pm 0.31$ & $\sim 4$\\
\hline
\end{tabular}
\flushleft
Note: \textsuperscript{a} Synthetic photometry in AB-magnitudes. \textsuperscript{b} WISE catalog photometry converted to AB-magnitudes. \textsuperscript{c} The difference between the synthetic and catalog AB-magnitudes with combined uncertainty.\textsuperscript{d} Number of standard deviations separating synthetic and catalog magnitudes.
\end{table*}

\section{Separating stars from background objects}
\label{sec:SED_modelling}

Despite the blended nature of our sources, careful modeling of the brighter-fatter effect (see below) allows us to attain relatively precise photometry for our sources in most of our filters. 

The Si:As impurity-band conduction detectors used by MIRI are subject to the brighter-fatter effect \citep{Argyriou23,Libralato24}, in which charge generated by a bright source laterally repels subsequent charge carriers, broadening the empirical PSF core relative to our model PSFs. That is to say, model PSFs will underestimate the core width of a bright source, producing a characteristic residual and biasing the measured flux. We provide an empirical correction for this by modelling the PSF core with a radial Voigt profile \citep{Argyriou23}, fit to the model PSF via least-squares minimization, yielding baseline Gaussian and Lorentzian width parameters $\sigma_*$ and $\gamma_*$. Each source, $i$, then receives its own Voigt profile, $V_i(r;\,\sigma_i\sigma_*\,\gamma_i\gamma_*$), with dimensionless free broadening scalings $\sigma_i$ and $\gamma_i$, where $\sigma_i=\gamma_i=1$ recovers the baseline Voigt profile and values greater than 1 broadens the core to absorb brighter-fatter excess. Thus, our brighter-fatter correction is $\Delta_i = V_i - V_*$, and the composite two-source model that we will be using for candidate D at F560W and F1000W (see \S~\ref{sec:SED_modelling}) is 

\begin{equation}
    \label{eq:bfcorr}
    M(\mathbf{x}) = f_1\!\left[P_1(\mathbf{x}) + \Delta_1(\mathbf{x})\right] + f_2\!\left[P_2(\mathbf{x}) + \Delta_2(\mathbf{x})\right] + b,
\end{equation}

\noindent where $P_i$ is the \texttt{stpsf} PSF shifted to source $i$'s centroid via cubic interpolation, $f_i$ is its flux, and $b$ is a constant background.

A full description of our methodology is provided in Appendix~\ref{app:bfc}, but here we briefly describe our analyses of candidates D and E. For candidate D, the M dwarf and background galaxy are separated enough that we can attempt to model their photometry directly. In the F560W and F1000W filters, both objects are relatively bright, so we fit both objects directly using the model presented Equation~\ref{eq:bfcorr}. Here, we sample 11 free parameters: centroid positions $(x_i, y_i)$, fluxes $f_i$, Voigt broadening scalings $(\sigma_i, \gamma_i)$, and a constant background $b$. We perform Bayesian nested sampling with $1,200$ live points using \texttt{dynesty}. Here, we placed flat priors of $\pm 3$~pixels on the centroid positions, flat priors of $[0,10^4]$~MJy~sr$^{-1}$ on the fluxes, and flat priors of $[0,10]$ on the broadening parameters. At F1500W, the M dwarf is too faint relative to the galaxy for a stable simultaneous fit to be achieved. We instead fit only the galaxy using a single-source variant of Equation~\ref{eq:bfcorr}, then recovered the stellar flux at its known position from the galaxy-subtracted residual image via circular aperture photometry within $r=0.5''$. Our photometry for candidate D at F1500W thus represents a lower limit as we are likely subtracting off some of the stellar flux with our model of the galaxy.

For candidate E, the M dwarf and galaxy are more severely blended, with the galaxy flux overwhelming the stellar PSF core in all three filters. However, we are still able to attain precise photometry in F560W using the MIRI diffraction spikes that extend well beyond the galaxy into relatively empty regions of the image. To do so, we first masked the western portion of the image (see, e.g., Figure~\ref{fig:MIRI_images}) as well as a disk of radius $1.1''$ that we centered on the the brightest pixel in the remaining half of the image. We then fit a single-source model to the unmasked region of the image, which, by construction, constitutes only the spikes that extend to the north, south, and east of the star. We first used the bright diffraction spikes in the masked image to set tight, informed priors on positions and fluxes. We use these priors to model the star in the unmasked image with \texttt{dynesty}. 
Since this geometry provides no leverage on the PSF core, we fixed the Voigt broadening parameters to those recovered from the candidate D F560W fit, a choice justified by the comparable fluxes of the two candidates at this wavelength (see, e.g., Table~\ref{table:photometry}). From there, we derive photometry for the galaxy by subtracting the model for the star off from the background-subtracted image and summing the flux remaining in a 3" aperture centered on the center of the galaxy.

In Table~\ref{table:photometry}, we list the total observed AB magnitudes of candidates D and E (i.e. star + background galaxy), and -- when possible -- the inferred AB magnitudes of the star and background galaxy separately.

\begin{table*}
\caption{Observed JWST/MIRI 560W, F1000W and F1500W AB magnitudes corresponding to the total observed flux (star + background galaxy),  and -- in cases where the two components can be separated -- the fluxes of star and galaxy separately.}
\label{table:photometry}
\begin{tabular}{cccccccccccc}
\hline
Target   & F560W$_\mathrm{tot}$ & F560W$_\star$  & F560W$_\mathrm{galaxy}$ & F1000W$_\mathrm{tot}$ 
& F1000W$_\star$ & F1000W$_\mathrm{galaxy}$ & F1500W$_\mathrm{tot}$ &  F1500W$_\star$ &  F1500W$_\mathrm{galaxy}$ \\ \hline
Candidate D  & $17.2\pm 0.1$      & $17.5\pm 0.1$  & $18.7\pm 0.1$  & $16.4\pm 0.1$   &  $18.5\pm 0.1$  & $16.6\pm 0.1$  &  $14.9\pm 0.1$      & $>19.8$ & $14.9\pm 0.1$ \\
Candidate E  & $17.0\pm 0.1$    & $17.2\pm 0.1$  & $18.7\pm0.1$  & $16.2\pm 0.1$   & --  & --   & $16.3\pm 0.1$      & --  & -- \\\hline
\end{tabular}
\end{table*}

\begin{figure*}
\includegraphics[width=\columnwidth]{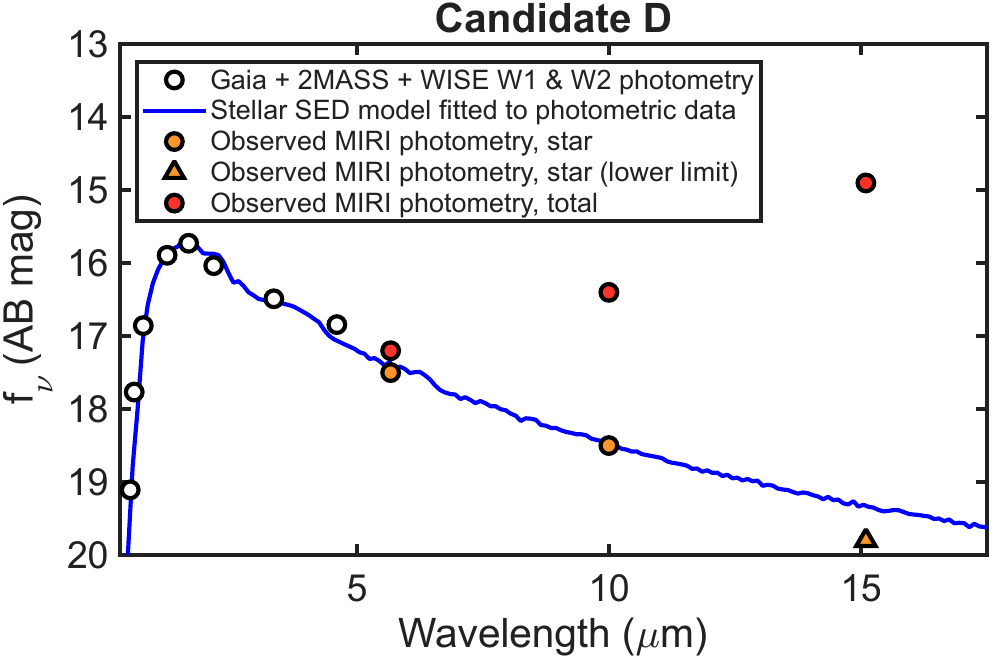}
\includegraphics[width=\columnwidth]{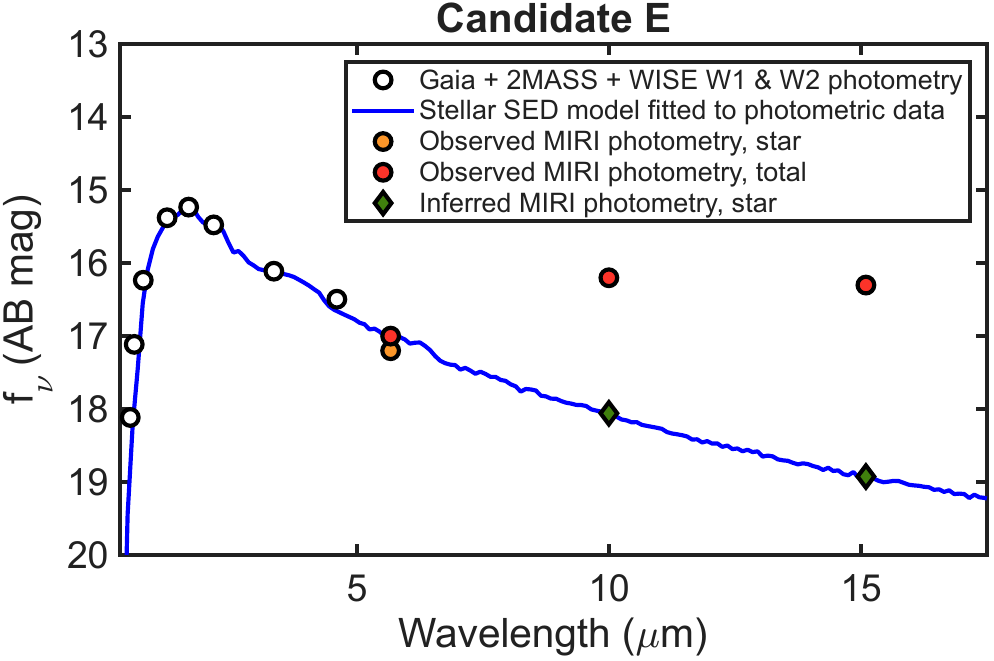}
\caption{Relative flux contributions from the M-dwarf stars to the total mid-IR flux of Project Hephaistos candidates D (left) and E (right). The blue lines represent rebinned Phoenix model stellar spectra scaled to approximately match the Gaia, 2MASS, and ALLWISE $W$1 and $W$2 photometry (white circles), in addition to the SPHEREx spectra (not shown, to avoid cluttering), at $\approx 0.5$--4.7 $\mu$m, where the M dwarf dominates the flux. Red circles represent the total MIRI F560W, F1000W and F1500W fluxes, and orange circles the fluxes of the M dwarfs in the MIRI filters where this was possible to measure directly. The photometric error bars are smaller than the symbols plotted. The orange triangle (left panel) represents a lower limit on the flux. Green squares (right panel) indicate the fluxes of the star inferred from Phoenix model fit in MIRI filters F1000W and F1500W, for which a direct measurement of the star failed due to blending with bright features within the background galaxy. The agreement between the measured MIRI fluxes of the M dwarfs fluxes and the continuum level of the stellar atmosphere model indicates that the M dwarfs have no significant intrinsic mid-IR excess in MIRI filters where direct flux measurements were possible. 
}
\label{fig:SED_fits}		
\end{figure*}

In Figure~\ref{fig:SED_fits}, we show Phoenix stellar atmosphere spectra \citep{Allard12} scaled to fit the Gaia DR3, 2MASS and ALLWISE W1 and W2 photometry ( $\approx 0.51$--4.6 $\mu$m) and the SPHEREx spectra (0.75--5.0 $\mu$m) of candidates D and E, i.e. at wavelengths for which the M dwarf models of \citet{Suazo24} were able to explain the observed SEDs without any IR excess. The Phoenix model spectra shown have $T_\mathrm{eff}=3500$ K, $\log (g)=5.0$ and [M/H]=0, similar to the  Gaia DR3  $T_\mathrm{eff}$ values (3473 K for candidate D; 3556 K for candidate E). These model spectra have been corrected for Milky Way extinction using the \citet{Fitzpatrick99} $R_V=3.1$ reddening curve and the Gaia DR3 $A_G$ values ($\approx 0.32$ mag for candidate D and $\approx 0.22$ mag for candidate E), and rebinned to lower resolution for plotting purposes. The Rayleigh-Jeans tails of these models are seen to be in rough agreement with the measured MIRI fluxes of the candidate D and E stars in the filters where it was possible to disentangle the light of the star from that of the background galaxy (F560W, F1000W and to some extent F1500W for candidate D; F560W for candidate E). Hence, there is no obvious evidence of any mid-IR excess associated with the stars themselves in our MIRI data. Under the assumption that there really is no intrinsic mid-IR excess associated with these M dwarfs, we may then use either the observed stellar fluxes, or the stellar continuum level inferred from the stellar atmosphere fit to estimate the relative contribution of the stars to the total observed mid-IR flux in the MIRI filters. For candidate D, the M dwarf is found to be dominant in the F560W filter ($\approx 80\%$ contribution), but subdominant in F1000W ($\approx 10\%$) and in F1500W ($\approx 2\%$  based on the continuum level of the stellar atmosphere model spectrum; $>1\%$ based on the measured MIRI lower limit on the flux). For candidate E, the corresponding values are $\approx 80\%$ in the F560W filter; $\approx 20\%$ in F1000W and $\approx 10\%$ in F1500W (with the latter two based on the continuum level of the stellar atmosphere model spectrum).

\section{Nature of the background objects}
\label{sec:nature of background objects}

\subsection{Spectral features and redshifts}
\label{subsec:features and z}
From the numerous emission features seen in the observed MIRI spectra (Figure~\ref{fig:MIRI_spectra _obsframe}), three emission lines were chosen for each candidate which allowed us to pin down the redshifts of the background objects to $z\approx 0.922 \pm 0.002$ (candidate D) and $z\approx 0.4104\pm 0.0003$ (candidate E). 
The chosen spectral lines are all fine-structure lines with low ionization potentials (IPs), that are less affected by AGN-driven blueshifts, with SNR $> 5$. Due to a lack of strong emission lines for candidate D, the [Fe II] emission line was also included after visual inspection. However, we excluded the [Fe II] emission line for candidate E due to its location between channels, with separate and conflicting detections of the spectral line in the two channels. We first determined the redshift for each spectral line independently, and then calculated the weighted average of them. The error is a conservative estimate, determined by the maximum difference in redshift between any pair of the three included spectral lines.

\begin{figure*}
\includegraphics[width=\columnwidth]{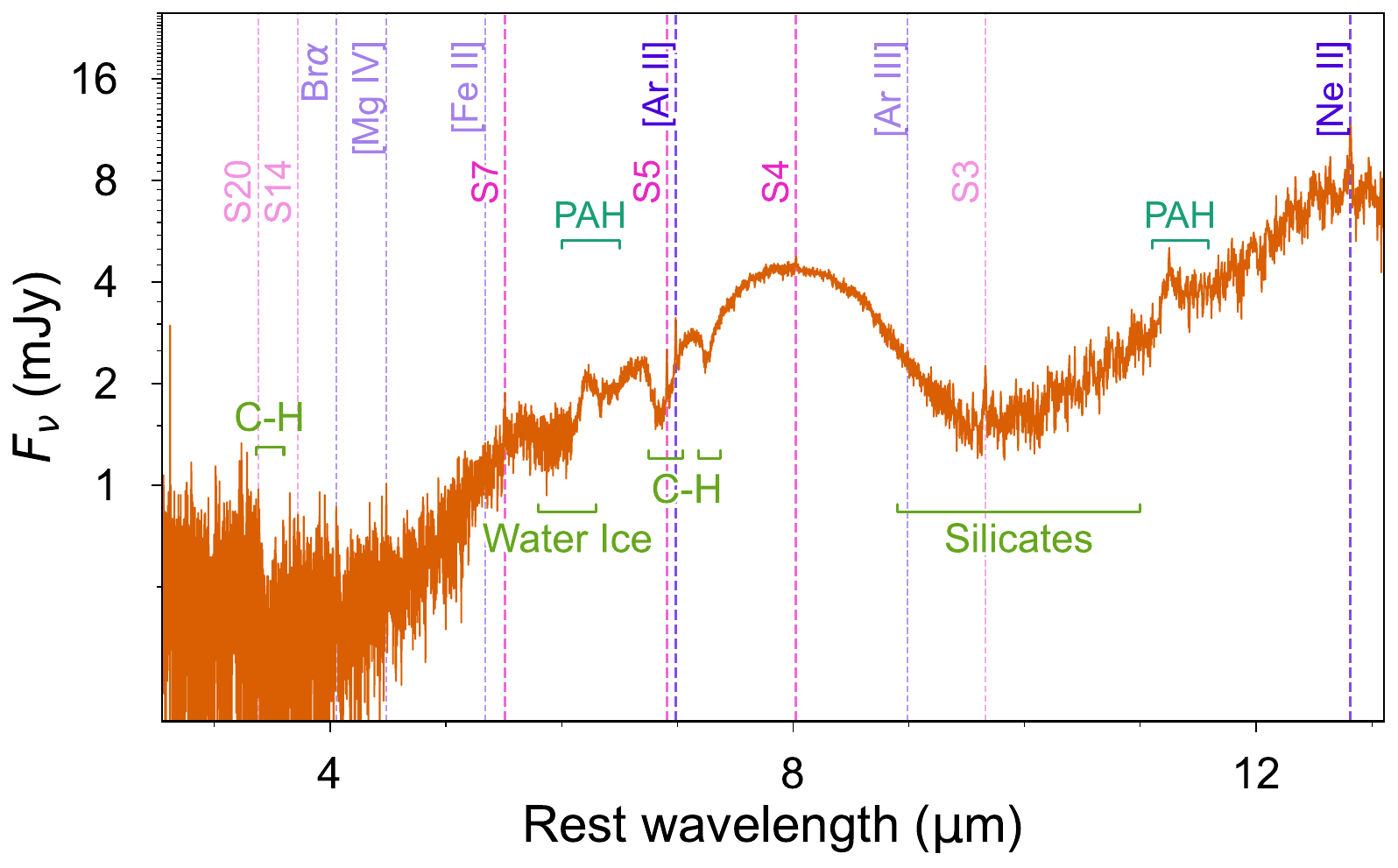}
\includegraphics[width=\columnwidth]{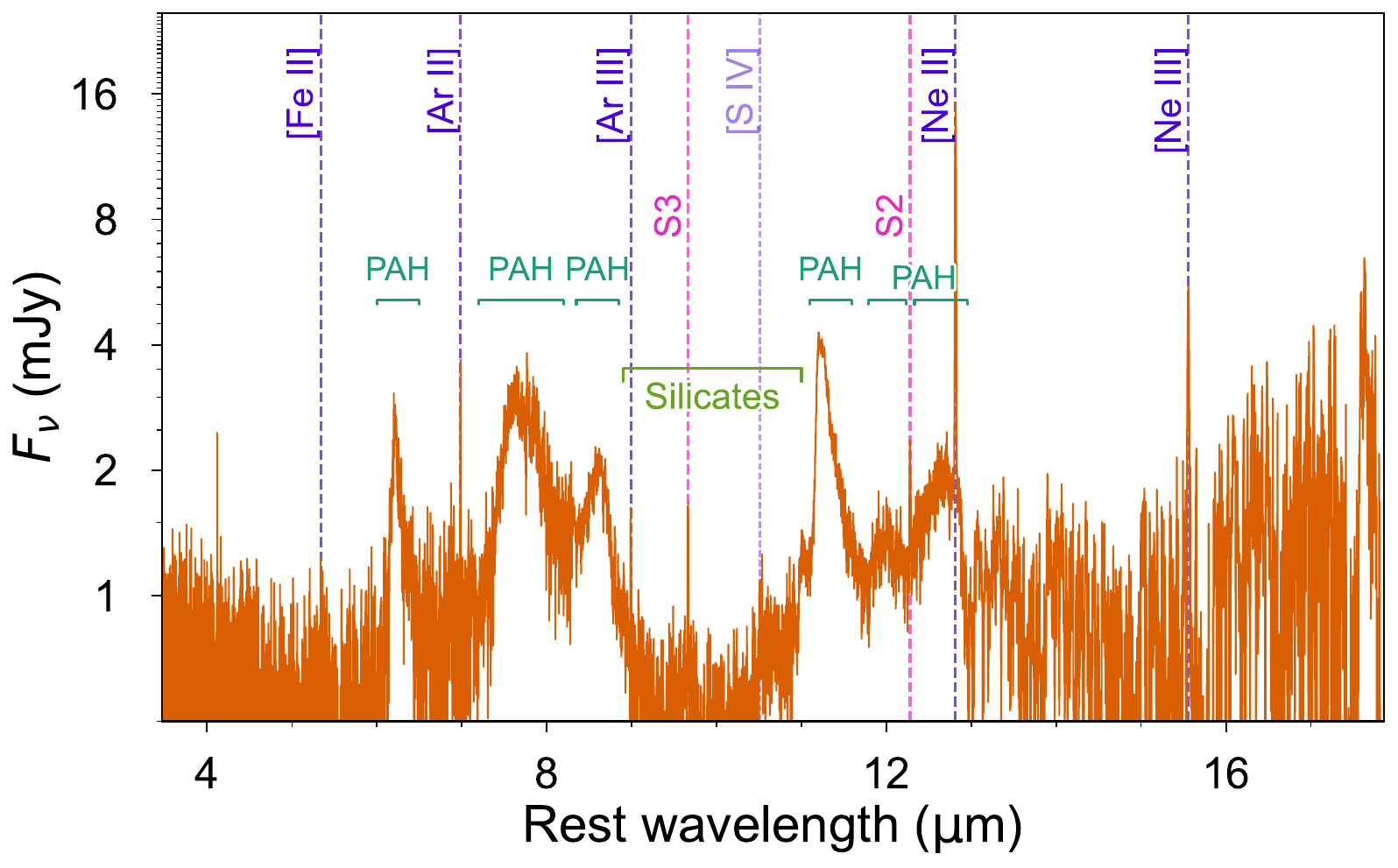}
\caption{Rest-frame MIRI spectra of candidate  D (left panel) and E (right panel) with spectral features identified based on the inferred redshifts of the background galaxies marked.}
\label{fig:MIRI_spectra_restframe}		
\end{figure*}

In Figure~\ref{fig:MIRI_spectra_restframe}, we show the corresponding redshift-corrected rest-frame spectra and the identified spectral features. This includes fine-structure atomic lines with low ($< 22$ eV) IPs and H$_2$ ro-vibrational lines. Candidate D shows a dominating, steeply rising (in F$_\nu$) continuum towards longer wavelengths which could indicate the presence of an AGN, weak PAH bands, and a moderately strong silicate absorption at $9.8 \upmu$m. It also shows aliphatic C-H and water-ice absorption bands. For candidate D, there is also a weak detection of the 4.5$\upmu$m [Mg IV] line which has a high IP (80 eV), evident of an AGN. However, due to the line's low SNR, we cannot rule out a noise artifact at its expected wavelength.

Candidate E, in contrast, has several strong PAH bands that dominate the spectrum, a weak silicate absorption and an underlying continuum that rises slightly towards longer wavelengths. Candidate E has one medium-high ionization potential line, the [Ne III] line, which could be created inside an AGN. However, [Ne III]/[Ne II] $= 0.31 \pm 0.02$ being lower than that of AGN-hosting galaxies suggests that the [Ne III] emission is entirely from massive, young stars \citep{PereiraSantaella2010}).

As seen in Figure \ref{fig:MIRI_images}, an unknown interloper was located northeast of the star in candidate D. A spectral extraction of solely this area showed little resemblance with the main background galaxy with no distinct spectral features, and did not match up with the galaxy's redshift. No redshift between $0 < z < 2$ could be determined for this unknown object.

Additionally, the spectrum over the star in candidate E was extracted separately to verify that a bright clump of the galaxy was the source of the mid-infrared emission. The spectrum showed the same emission features as that of the star + background galaxy of candidate E, verifying that part of the galaxy must indeed overlap the star. This validates the conclusion that there is no mid-IR excess associated with the stars themselves.

Note that we in Figure~\ref{fig:MIRI_spectra_restframe} apply the same redshift correction across the whole spectrum, even though a substantial fraction of the flux at the shorter-wavelength part of this spectrum comes from the foreground M dwarfs (Section~\ref{sec:SED_modelling}). While this affects the inferred rest-frame continuum level of the background sources, this has little effect on the identification of spectral features or the inferred redshift, since these M dwarfs are not expected (nor are observed) to display strong emission features in this wavelength regime.

\subsection{Photometric SED analysis}
\label{subsec: SED analysis}
\citet{Ren24} and \citet{Blain24} propose that the mid-IR WISE excess of the \citet{Suazo24} candidates may be caused by source confusion with background Hot DOGs, characterized by very steep (red-tilted) SEDs in the rest-frame $\approx 1$--10 $\mu$m range \citep{Tsai15,Assef15,Ricci17}. This hypothesis can be tested for candidates D and E, at $z\approx 0.9$ and $z\approx 0.4$ respectively, by comparing the SED of the background galaxy to those of Hot DOGs as well as other classes of IR-bright objects. For this exercise, we make use of the shape of the observed SED for which the brightness of the background galaxy can be separated from that of the M dwarf (Table~\ref{table:photometry}), or for which the brightness of the M dwarf is deemed subdominant (observed wavelength $\gg 5\ \mu$m; Figure~\ref{fig:SED_fits}). We also factor in limits on the bolometric luminosity from non-detections in the far-IR, and that potential SEDs with optical-to-mid-IR flux ratios above some limit are ruled out because they would make the background object visible also in the optical.

\subsubsection{Luminosity constraints}
While the Hot DOGs identified by \citet{Tsai15} and \citet{Assef15} at  $z\approx 2$--4 can be classifed as 
Extremely Luminous Infrared galaxies (ELIRGs; bolometric luminosity $L_\mathrm{bol}>10^{14}\  L_\odot$) or Hyper-Luminous Infrared Galaxies (HyLIRGs; $L_\mathrm{bol}=10^{13}$--$10^{14}\  L_\odot$), the $z\lesssim 1$ Hot DOGs studied by \citet{Li25} fall in the range of Luminous Infrared Galaxis (LIRGs; $L_\mathrm{bol}=10^{11}$--$10^{12}\  L_\odot$) or Ultra-luminous Infrared Galaxies (ULIRGs; $L_\mathrm{bol}=10^{12}$--$10^{13}\  L_\odot$). Even though we lack the far-infrared detections of the background galaxies of candidate D and E that would be required to measure their total IR luminosities, the non-detections of these objects in the  IRAS Point Source Catalog v2.1 \citep{IRAS_point_source_catalog} indicate that the brightness at 25, 60 and 100 $\mu$m must be $<0.5$, $<0.6$ and $<1.0$ Jy, respectively.  By adding a blackbody component, at the redshift of the background galaxies, that connects to the observed-frame WISE $W$4 fluxes at 22 $\mu$m and contributes maximally at longer wavelengths without overshooting the IRAS limits, and integrating this curve out to a rest-frame wavelength of 1000 $\mu$m, we can derive the maximally allowed IR luminosity (which, for these IR-dominated SEDs, also serves as a proxy for the bolometric luminosity $L_\mathrm{bol}$). The resulting upper limits are $L_\mathrm{bol}<5\times 10^{13}\ L_\odot$ for the background galaxy of candidate D (limit met for a blackbody temperature of  $\approx 90$ K) and $L_\mathrm{bol}<7\times 10^{12}\ L_\odot$ for the background galaxy of candidate E (limit met for a blackbody temperature of  $\approx 70$ K). Hence, the candidate D galaxy could potenitally have a bolometric luminosity as high as in the HyLIRG regime (but also much lower), whereas the candidate E galaxy is limited to the luminosity regime of ULIRGs (or lower). Neither limit contradicts the Hot DOG hypothesis. 

\subsubsection{SED shape constraints}
\begin{figure*}
\includegraphics[width=\columnwidth]{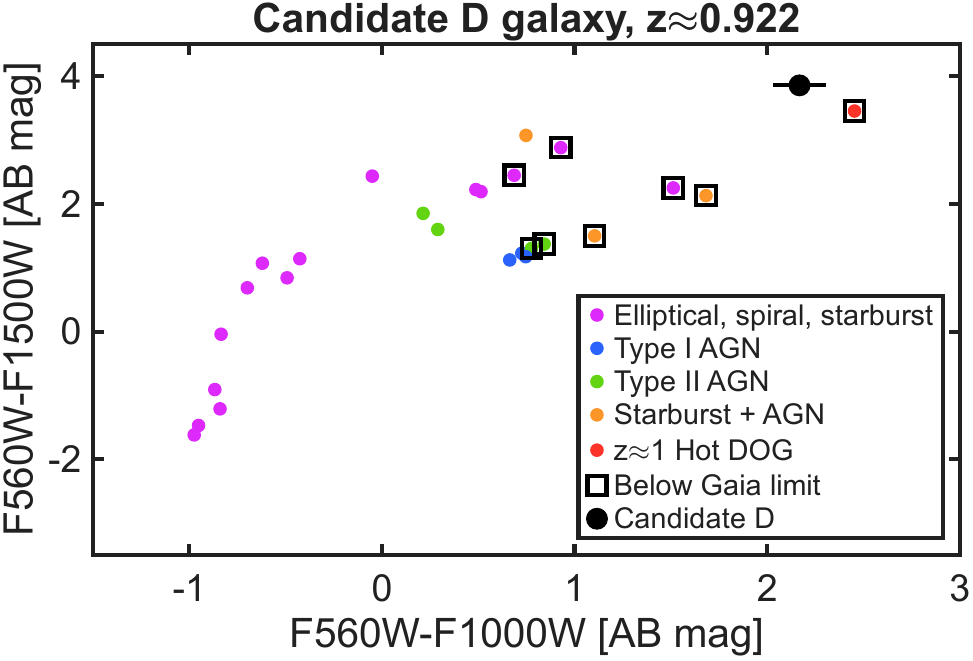}
\includegraphics[width=\columnwidth]{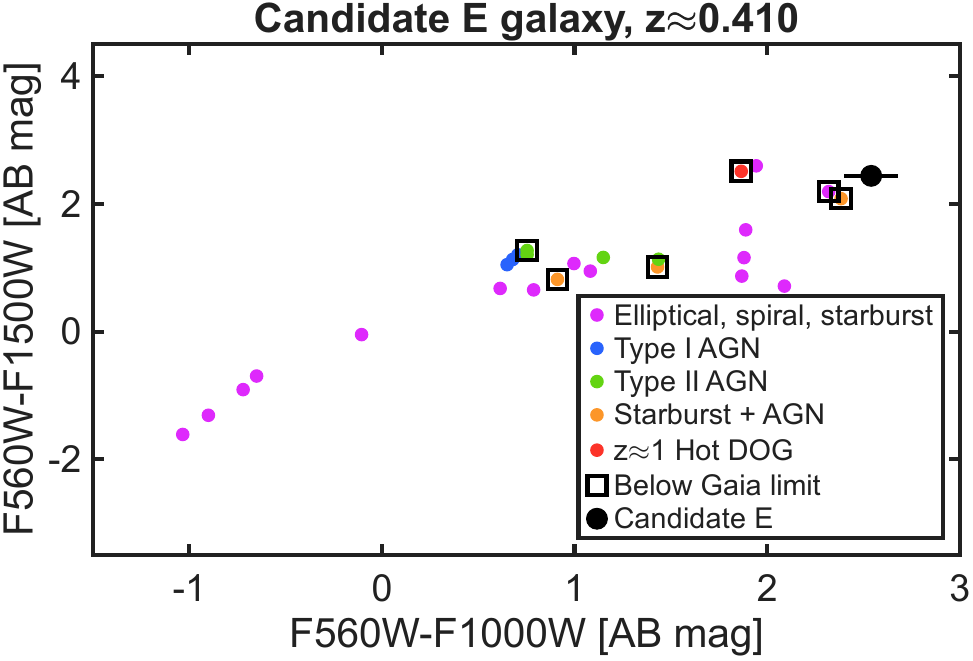}
\caption{Observed MIRI F560W--F1000W vs. F560W--F1500W colours of the candidate D (left panel) and candidate E (right panel) background galaxies, compared to the corresponding colours derived from template SEDs of IR-bright objects redshifted to $z=0.922$ and $z=0.410$ respectively. The large black circles with error bars (smaller or comparable to the circle in F560W--F1500W) mark the observed fluxes of the candidate D and E background galaxies and the red circle the \citet{Ricci17} $z\approx 1$ Hot DOG template. Additional circles represent templates from the \citet{Polletta07} set for elliptical, spiral and starburst galaxies (cyan), type I AGNs (blue), type II AGNs (green) and starbursts+AGNs (orange). Templates which, when scaled to the F1000W flux of the candidate D and E galaxies, would potentially remain undetected in the optical (in accordance with the lack of objects close to the candidate D and E M-dwarf stars in Gaia data) are marked by black squares. While the MIRI colours of the candidate D galaxy most closely resembles the \citet{Ricci17} $z\approx 1$ Hot DOG, the MIRI colours of the candidate E galaxy are also consistent with several other types of objects, including the starburst galaxy NGC 6090.}
\label{fig:MIRI_col_col_diagram}
\end{figure*}

The Hot DOG hypothesis can be further tested by both comparing the shape of the photometric SEDs of the candidate D and E background galaxies across the MIRI F560W, F1000W and F1500W filters to the corresponding SED shapes of various SED templates for infrared-bright objects. In the case of the candidate D and E galaxies, these filters probe the SED at rest-frame wavelengths $\approx 2.6$--8.6 $\mu$m and $\approx 3.6$--11.7 $\mu$m, respectively. In Figure~\ref{fig:MIRI_col_col_diagram}, we compare the F560W--F1000W vs. F560W--F1500W colours of the candidate D and E galaxies to the corresponding colours derived from the \citet{Ricci17} $z\approx 1$ Hot DOG SED  and the \citet{Polletta07}  SWIRE template library for elliptical, spiral and starburst galaxies, type I active galactic nuclei (AGN), type II AGNs and objects labeled as AGN with prominent starbursts. Since the number of wavelength data points of the \citet{Ricci17} $z\approx 1$ Hot DOG SED is small (10 data points in the rest-frame $\approx 0.18$--170 $\mu$m range), this SED has been interpolated linearly in rest-frame wavelength and $f_\lambda$ flux before synthetic MIRI photometry was derived. All of the comparison SEDs have been redshifted to match the candidate D and E galaxies. While many of these templates display very red MIRI colours, some of them have too high optical-to-IR flux ratios to match the overall SED characteristics of our candidates. Since the candidate D and E background galaxies are undetected in Gaia DR3, we discount templates that -- when scaled to the F1000W fluxes of D and E -- would be brighter than 20 AB mag in the Gaia G-band. Template SEDs that are sufficiently red-tilted to evade detection by Gaia based on this constraint, and hence remain viable models for our targets, are marked by black squares in Figure~\ref{fig:MIRI_col_col_diagram}.

For the candidate D galaxy, the MIRI colours resemble the \citet{Ricci17} Hot DOG SED more than any of the \citet{Polletta07} SEDs, and is redder by $\approx 0.5$ mag compared to their reddest F560W--F1000W object (starburst+AGN IRAS 19254-7245 South template). Hence, the measured MIRI colours indicate that the background galaxy of candidate D may well be a Hot DOG, and this is also favoured by its point-like appearance (Figure~\ref{fig:MIRI_images}), which indicates a dominant AGN. 

The situation is less clear-cut for the background galaxy of candidate E. Although its F560W--F1000W and F560W--F1500W colours are both very red, with the F560W--F1500W colour just $\approx 0.1$ mag away from that derived from the \citet{Ricci17} Hot DOG SED, a few of the \citet{Polletta07}  objects occupy a similar region in this MIRI colour-colour diagram (the starburst NGC 6090 and the starburst+AGN IRAS 22491-1808, all with F560W-F1500W $\gtrsim$ 2 mag) due to prominent PAH features at rest-frame wavelengths $\approx 6.2$, 7.7 and 11.3 $\mu$m in the F1000W and F1500W filters. 
Hence, the SED of candidate E is broadly consistent with that of a dusty starburst, with or without an AGN. 

\subsection{Spectral analysis}
In order to describe the nature of the background galaxies based on the spectral features, several diagnostic tools were used. The primary tool is the Spoon diagram developed by \citet{Spoon07}, where the strength of the 10 $\upmu$m silicate absorption feature ($S_{sil}$) is plotted against the equivalent width (EW) of the $6.2 \upmu$m PAH band. Depending on their location in this diagram, galaxies are classified into one of 9 classes ranging from starburst dominated spectra (1C), to deeply obscured nuclei (3A), and power-law dominated AGNs (1A).

\begin{figure}
    \centering
    \includegraphics[width=\columnwidth]{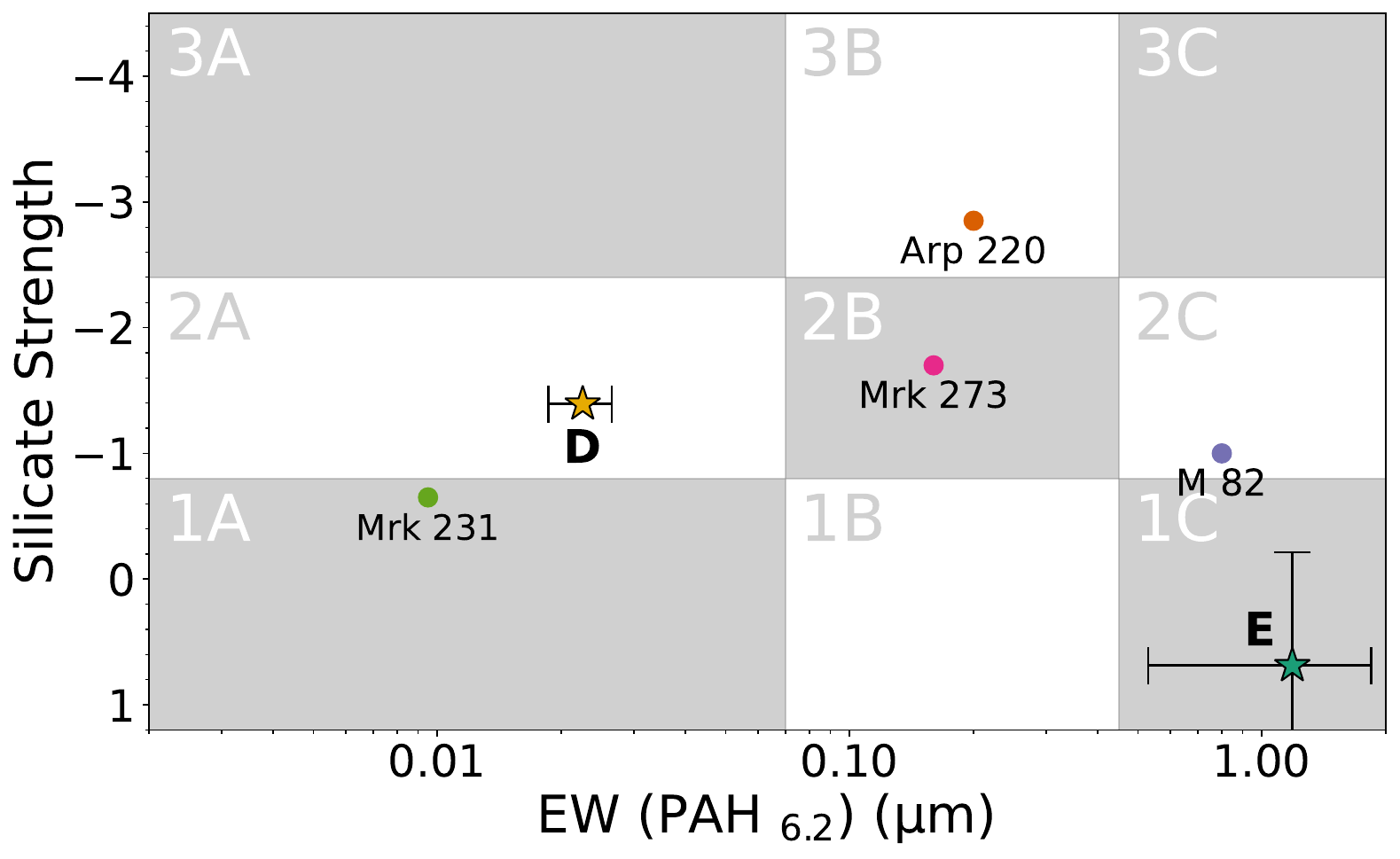}
    \vspace{-0.4cm}
    \caption{Candidates D (2A: obscured AGN) and E (1C: starburst) plotted in the Spoon diagram, together with several well studied galaxies. Note that candidate E stays firmly within the 1C class, despite the large error bars. Mrk 231 (1A) is a local ULIRG and quasar-host, Mrk 273 (2B) and Arp 220 (3B) are ULIRGs in the midst of a merger, and M 82 (2C) is a starburst galaxy.}
    \label{fig:spoondiagram}
\end{figure}

We used a slightly modified version of the procedure presented in \citet{Spoon07} in order to determine EW(PAH$_{6.2}$) and $S_{sil}$. The main modification consisted of accounting for the strong noise at longer wavelengths. In practice, for $S_{sil}$, we modeled the underlying continua with a spline interpolation with pivot points at 5.2, 5.6, 7.8, and 13$\upmu$m (candidate D) and with a power-law interpolation from 5.5 to 14.5$\upmu$m (candidate E). For EW(PAH$_{6.2}$), we calculated the integrated flux with a continuum based on a linear interpolation between the anchor points 5.45 and 7.05$\upmu$m. This procedure was applied with the Phoenix model spectra subtracted from the total spectra, the results are presented in Table \ref{tab:spoon_table}. Based on these results, candidate D fits into the 2A class of obscured AGNs, while candidate E belongs to the 1C starburst-dominated class, as shown in Figure \ref{fig:spoondiagram}. 

\begin{table}
\centering
\caption{Calculated quantities used in the Spoon diagram (Figure \ref{fig:spoondiagram}). The given uncertainties are very conservative, and were determined by taking the difference between the values calculated using the total spectrum and those with the star-subtracted spectrum. The weak and noisy continuum of the galaxy in candidate E, combined with a lower redshift and higher stellar contribution, makes it very sensitive to changes in initial condition, i.e. slight variations in the Phoenix stellar model.}
\label{tab:spoon_table}
\begin{tabular}{lrr}
\hline
Candidate & EW(PAH$_{6.2}$) [$\upmu$m] & $S_{sil}$ \\
\hline
D & $ 0.023\pm 0.004$ & $ -1.41 \pm 0.04$\\
E & $ 1.2 \pm 0.7$ & $0.7 \pm 0.9$\\
\hline
\end{tabular}
\end{table}

Moreover, the \texttt{CAFE} spectral decomposition package \citep{CAFE} was used, again with the stars' model spectra subtracted from the total spectra, to determine a dominant dust component for the continua with temperatures $(92 \pm 10)$ K (candidate D) and $(67 \pm 4)$ K (candidate E). The errors were achieved through Monte Carlo iteration. Furthermore, we obtained the intrinsic PAH strength also with the \texttt{CAFE} package and used it to determine lower bounds for the star-formation-rates (SFRs) of $\sim   70\ M_\odot \text{yr}^{-1}$ (candidate D) and $\sim    20\ M_\odot \text{yr}^{-1}$ (candidate E). This was done using the $L_{\rm PAH}$-SFR-relations in \citet{Shipley2016}, which are only calibrated up to values of $100 M_\odot \text{yr}^{-1}$. The results imply that both galaxies are actively star-forming, with candidate D in particular entering the SFR regime of ULIRGs \citep{daCunha2010}.

In summary, we find that the galaxy of candidate D is quite Hot DOG-like. This is due to its steeply increasing (in F$_\nu$) continuum toward longer wavelengths, its placement in class 2A on the Spoon diagram, and the dust temperature being comparable to that of WISE-selected Hot DOGs \citep{Fan2016}. Although the lower limit of the SFR is two orders of magnitude smaller than the SFRs of WISE-selected Hot DOGs \citep{Fan2016}, this is not unexpected, due to the destruction of PAH molecules via a strong AGN. Secondly, the galaxy of candidate E shows several features inconsistent with Hot DOGs. The spectrum is clearly starburst-dominated, with strong PAH-bands and a weak continuum. This is verified by its placement in the 1C class in the Spoon diagram. The lack of high IP spectral lines also fail to indicate the presence of an AGN. Overall, many factors defy the Hot DOG-hypothesis. There is, however, one category of DOGs that line up with these characteristics: the ``bump'' DOGs first studied by \citet{Dey2008}. It is still difficult to verify this classification, as the tell-tale sign of these ``bump'' DOGs is the $1.6 \upmu$m bump in the spectrum, which in our case would be fully over-powered by the foreground star. The dust temperature retrieved from \texttt{CAFE}, however, are more consistent with Hot DOGs (60-120 K) than other DOGs (20-40 K) \citep{Wu12}. 

\section{Discussion}
\label{sec:discussion}
The JWST/MIRI imaging alone rules out both the Dyson-sphere  and extreme debris-disk explanations for the mid-IR excess of candidates D and E, since both of these scenarios would predict the mid-IR radiation to originate at distances too close to the M dwarfs to be resolvable in MIRI imaging. A Dyson sphere or an extreme debris disk that radiates as a blackbody with $T\approx 180$ K and $L_\mathrm{IR}/L_\mathrm{tot}\approx 0.1$  \citep[as inferred by][for candidates D and E]{Suazo24} would need to be located at $\lesssim 1$ AU from these $T_\mathrm{eff}\approx 3500$ K, $L_\mathrm{tot}\lesssim 0.03\ L_\odot$  M dwarfs, which at distances of $\gtrsim 200$ pc would correspond to an angular diameter of $\lesssim 0.01$ arcsec, i.e. more than an order of magnitude smaller than the resolution limit of MIRI. Since a secondary point-source (candidate D) or an extended object (candidate E) are clearly seen at distances $\sim 1$ arcsec from the M dwarfs (Figure~\ref{fig:MIRI_images}), background objects are favoured as the source of the mid-IR excess, and this conclusion is further enforced by the MIRI spectra, where redshifted emission features can be identified that place the mid-IR sources at cosmological distances (Figure~\ref{fig:MIRI_spectra_restframe}). Here, we discuss the consequences of these findings for future searches for Dyson spheres and for non-technosignature astronomy.

\subsection{Estimates of the contamination probability}
\citet{Suazo24} argue, based on estimated contamination rates, that only a minor fraction of their Dyson-sphere candidates are likely to have their mid-IR fluxes affected by projection close to background objects with very red SEDs. Follow-up studies nonetheless indicate that many of the candidates do in fact seem to be contaminated based on radio detections \citep{Ren24,Ren25}, or deeper near-IR data and optical-to-IR centroid shifts \citep{Ren26}. As shown in the current paper, at least two of the candidates (D and E) clearly have IR-bright background galaxies with projected distances as small as $\approx 1$ arcsec from the M dwarf (Figure~\ref{fig:MIRI_images}). So, what went wrong in the \citet{Suazo24} argument? 

\citet{Suazo24} estimate that the surface density of WISE unresolved galaxies with sufficiently bright $W$3 and $W$4 fluxes to contaminate the mid-IR fluxes in their sample of stars is $\approx 15000$ sr$^{-1}$. This estimate also includes a criterion on the $W3-W4$ colours of these galaxies, required to make them similar to their Dyson-sphere models across these two filters. Within a circular area with radius 3.25 arcsec (their example, i.e. inside WISE $W$3 point-spread function) or $\approx 1.0$ arcsec (similar to the projection distance of the candidate D and E galaxies) from a given star, one then expects a background galaxy in a fraction $1.2\times 10^{-5}$ or  $1.1\times 10^{-6}$ of the cases. \citet{Suazo24} then argue that this fraction cannot be applied to the full sample of $5\times 10^6$ stars, but only to the $2\times 10^5$ object subsample with WISE $W$3 and $W$4 detections. This reasoning seems risky, since it assumes that all of the $2\times 10^5$ stars with $W$3 and $W$4 detections in the catalog entered the sample due to having sufficiently high {\it intrinsic} fluxes in these filters, whereas in reality some fraction of these could have been lifted above the WISE $W$3 and $W$4 detection threshold precisely because of source confusion with an IR-bright background galaxy. If one were to instead apply the $1.2\times 10^{-5}$ (separation $\leq 3.25$ arcsec) or $1.1\times 10^{-6}$ (separation $\leq 1$ arcsec) contamination fractions to the full sample of $5\times 10^6$ stars, then one would expect $\approx 60$ or $\approx 6$ such contamination cases, respectively. However, this does not represent a reasonable estimate of the actual contamination rate in the \citet{Suazo24} selection pipeline either, because of the way they derive the surface number density of contaminating background galaxies ($\approx 15000$ sr$^{-1}$). The WISE colour critera used for these background galaxies include a wide range of W4-bright objects, including those that likely would be rejected by other parts of the selection pipeline due to causing conspicuous contamination in optical and/or near-IR data – either as a close companion seen in imaging, or by causing a compound photometric SED that would be inconsistent with their compound SED models (star plus Dyson sphere). A more realistic contamination estimate should be based on the surface number density of  sources with very red SEDs from the optical to the mid-IR that appear unresolved in WISE imaging. Hot DOGs do have SED shapes of this type, as pointed out by \citet{Ren24,Blain24,Ren25}, but are their number densities sufficient? If these objects are responsible for all the \citet{Suazo24} candidates, then we would require these objects to exhibit a surface number density in the range $\geq 0.5$ deg$^{-2}$ (7 Hot DOGs within 3.25 arcsec of a star in a sample of $5\times 10^6$ stars) or $\geq 6$  deg$^{-2}$ (7 Hot DOGs within 3.25 arcsec of a star in a sample of $5\times 10^6$ stars). \citet{Assef15} estimate 0.032 deg $^{-2}$ for $z=2$--4 Hot DOGs, and \citet{Li25} 0.0024 deg$^{-2}$ for $z<0.5$ Hot DOGs. While these numbers fall short by several orders of magnitude, it is important to realize that they are derived for Hot DOGs that are significantly brighter in WISE bands than the majority of the \citet{Suazo24} candidates, which typically lie  $\approx 1$ mag below the selection limit of the \citet{Assef15} sample in WISE $W$3 (only candidate A lies above) and are $\approx 3$--5 mag fainter in $W$3 than the \citet{Li25} $z<0.5$ Hot DOGs. Attempting to estimate the Hot DOG surface number densities at significantly fainter limits than done in the \citet{Assef15} and \citet{Li25} samples (and beyond the 95\% completion limit of the ALLWISE catalog) is beyond the scope of this paper, but given that it is not unreasonable for the number counts to increase by one order of magnitude when going 1 magnitude fainter, it is entirely possible that Hot DOG-like objects, at brightness levels fainter than typically targeted by Hot DOG studies, explain the mid-IR excess of all the \citet{Suazo24} candidates.

\subsection{Importance for future Dyson sphere searches}
The fact that a substantial fraction of the Project Hepaistos Dyson-sphere candidates are contaminated by background galaxies at arcsecond-scale separations (\citealt{Ren24,Ren25,Ren26}; this paper) has important consequences for future searches for Dyson spheres based on mid-IR ($\gtrsim 5\mu$m) excess. The fundamental issue lies in the limited angular resolution of existing all-sky mid-IR imaging databases (FWHM $\approx 6$ arcsec in the WISE $W$3 filter at 12 $\mu$m; $\approx 12$ arcsec in the WISE $W$4 filter at $\approx 24$ $\mu$m), which allows for substantial source confusion when optical, near-IR and mid-IR imaging is matched up. Although there are proposed mid-IR telescopes which in the 2030s could provide higher-resolution data over wide areas, e.g. PRIMA  \citep[The PRobe far-Infrared Mission for Astrophysics; FWHM$\approx 5$ arcsec at 24--45 $\mu$m;][]{Glenn25}, with the potential to survey $\sim1/4$ of the sky \citep{Burgarella25}, and GREX-PLUS \citep[Galaxy Reionization EXplorer and PLanetary Universe Spectrometer; FWHM $\approx 1$--2 arcsec resolution at 5--8 $\mu$m;][]{Inoue23}, near-future improvements must lie at the data-analysis stage. Alternative selection strategies based on machine learning have been proposed \citep{Contardo24,Mignone26}, but it may also be useful to include centroid offsets (which can typically be measured at smaller scales than the formal FWHM of the point spread function) between the higher-resolution optical/near-IR and the lower-resolution mid-IR data \citep{Ren26} as a quality criterion early in the selection pipeline for Dyson-sphere candidates to weed out interlopers in the form of background galaxies. It should be noted, however, that this method provided no evidence for interloper contamination in the case of candidate D and only  marginal evidence in the case of candidate E.

 \subsection{Natural guide stars for adaptive optics}
 In general, searches for Dysonian waste-heat signatures in large datasets will tend to single out astronomical objects with highly unusual optical-to-infrared properties. Follow-up observations of such cases could lead to discoveries of rare astrophysical phenomena, but also data on classes of objects which otherwise would not have been selected for detailed studies. The low-redshift ($z\lesssim 1$), red and IR-bright galaxies uncovered in the present work are significantly fainter than most well-studies Hot DOGs, which may make them interesting in their own right. What truly make these targets stand out \citep[in a sample of $5\times 10^6$ stars;][]{Suazo24}, however, is the small angular separation between the M dwarfs and these background IR-bright sources. This configuration could allow for high-resolution imaging and spectroscopy of the background galaxies using the foreground M dwarfs as natural guide stars in adaptive-optics observations. While the M dwarfs studies in this work ($\approx 14$ Vega mag in the K band) are a bit fainter than optimal for Single Conjugate Adaptive Optics (SCAO) observations at $\approx 3$--13 $\mu$m using the Mid-infrared ELT Imager and Spectrograph (METIS) on the Extremely Large Telecope \citep{Brandl21,Feldt26}, it may be worthwhile to explore the viability of such measurements in future work.

\section{Summary}
\label{sec:summary} 
Based on JWST/MIRI imaging and MRS data, we attribute the mid-IR excess of the M-dwarf, Dyson-sphere candidates D and E from \citet{Suazo24} to IR-bright background galaxies at redshifts $z\approx 0.9$ and $z\approx 0.4$, which lie at projected distances  $\approx 1$ arcsec from the M dwarfs. We find no evidence of excess mid-IR flux coming from regions at AU-scale distances from the M dwarfs themselves, which would be the case if the mid-IR excess flux had originated from a Dyson sphere or extreme debris disk. MIRI photometry of the background galaxies suggest that they have Hot DOG-like spectral energy distributions (SEDs) at observed wavelengths $\approx 5$--15 $\mu$m. Both the morphological appearance (point-like) and the MIRI spectrum of candidate D indicates that this object features an AGN. The more extended, clumpy morphology of candidate E, along with a MIRI spectrum more similar to those of dusty starburst galaxies indicates that any AGN must be subdominant, if at all present. A spectral decomposition of the MIRI spectra does reveal a hot dust component ($\approx 90$ and $\approx 70$ K) for both galaxies, which would support a Hot DOG interpretation, but if so, they have fainter mid-IR fluxes than currently identified Hot DOGs at $z<1$. 

\section*{Acknowledgements}
This work is based on observations made with the NASA/ESA/CSA James Webb Space Telescope. The data were obtained from the Mikulski Archive for Space Telescopes at the Space Telescope Science Institute, which is operated by the Association of Universities for Research in Astronomy, Inc., under NASA contract NAS 5-03127 for JWST. These observations are associated with program 7199. Support for program 7199 was provided by NASA through a grant from the Space Telescope Science Institute, which is operated by the Association of Universities for Research in Astronomy, Inc., under NASA contract NAS 5-03127. This research has made use of the NASA/IPAC Infrared Science Archive, which is funded by the National Aeronautics and Space Administration and operated by the California Institute of Technology.

This work has made use of data from the European Space Agency (ESA) mission
{\it Gaia} (\url{https://www.cosmos.esa.int/gaia}), processed by the {\it Gaia}
Data Processing and Analysis Consortium (DPAC,
\url{https://www.cosmos.esa.int/web/gaia/dpac/consortium}). Funding for the DPAC
has been provided by national institutions, in particular the institutions
participating in the {\it Gaia} Multilateral Agreement.

The Pan-STARRS1 Surveys (PS1) and the PS1 public science archive have been made possible through contributions by the Institute for Astronomy, the University of Hawaii, the Pan-STARRS Project Office, the Max-Planck Society and its participating institutes, the Max Planck Institute for Astronomy, Heidelberg and the Max Planck Institute for Extraterrestrial Physics, Garching, The Johns Hopkins University, Durham University, the University of Edinburgh, the Queen's University Belfast, the Harvard-Smithsonian Center for Astrophysics, the Las Cumbres Observatory Global Telescope Network Incorporated, the National Central University of Taiwan, the Space Telescope Science Institute, the National Aeronautics and Space Administration under Grant No. NNX08AR22G issued through the Planetary Science Division of the NASA Science Mission Directorate, the National Science Foundation Grant No. AST-1238877, the University of Maryland, Eotvos Lorand University (ELTE), the Los Alamos National Laboratory, and the Gordon and Betty Moore Foundation.

This publication makes use of data products from the Two Micron All Sky Survey, which is a joint project of the University of Massachusetts and the Infrared Processing and Analysis Center/California Institute of Technology, funded by the National Aeronautics and Space Administration and the National Science Foundation.

This publication makes use of data products from the Wide-field Infrared Survey Explorer, which is a joint project of the University of California, Los Angeles, and the Jet Propulsion Laboratory/California Institute of Technology, funded by the National Aeronautics and Space Administration.

This publication makes use of data products from the Spectro-Photometer for the History of the Universe, Epoch of Reionization and Ices Explorer (SPHEREx), which is a joint project of the Jet Propulsion Laboratory and the California Institute of Technology, and is funded by the National Aeronautics and Space Administration.

AB was supported by the Swedish National Space Agency.  RJA was supported by FONDECYT grant number 1231718 and by the ANID BASAL project FB210003. ORCID: 0000-0002-9508-3667.

\section*{Data Availability}
The JWST data that support the findings of this study are currently subject to an exclusive access (proprietary) period under Program ID GO 7199. Upon expiration of this period, the raw and calibrated data will become publicly available via Mikulski Archive for Space Telescopes (MAST).



\bibliographystyle{mnras}
\bibliography{references} 




\appendix
\section{Demonstration of the empirical brighter-fatter correction and the F560W diffraction-spike fit}
\label{app:bfc}

\begin{figure*}
    \centering
    \includegraphics[width=0.8\textwidth]{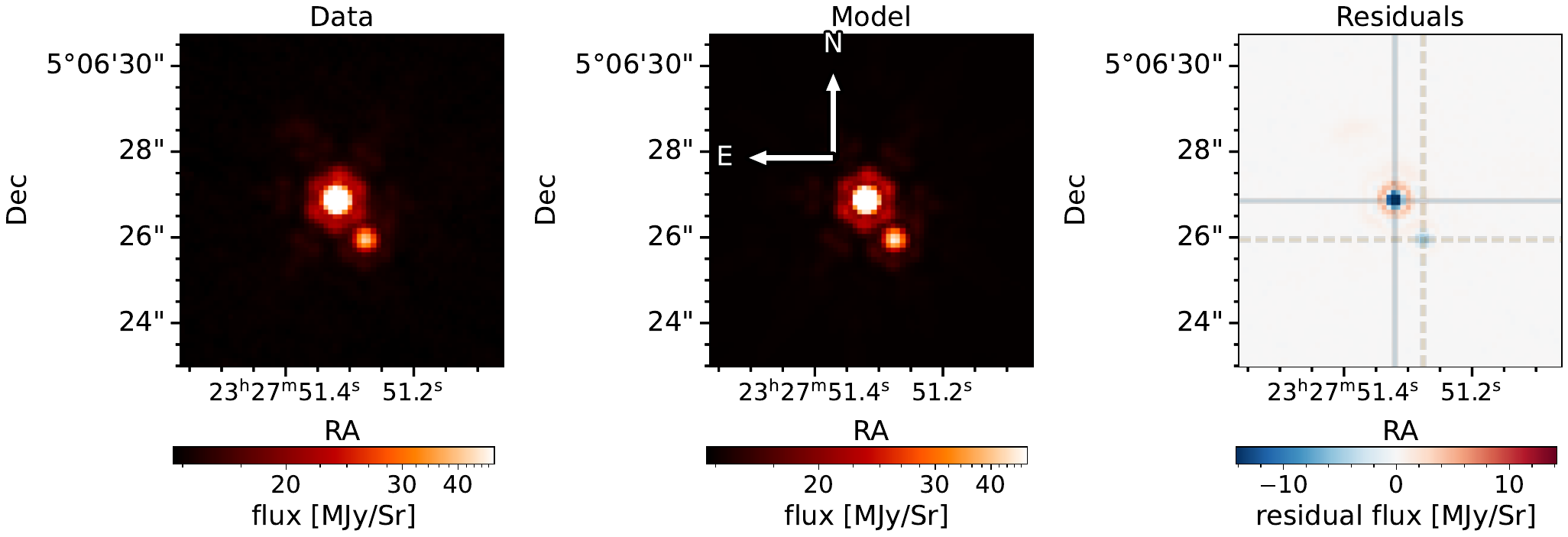}
    \includegraphics[width=0.8\textwidth]{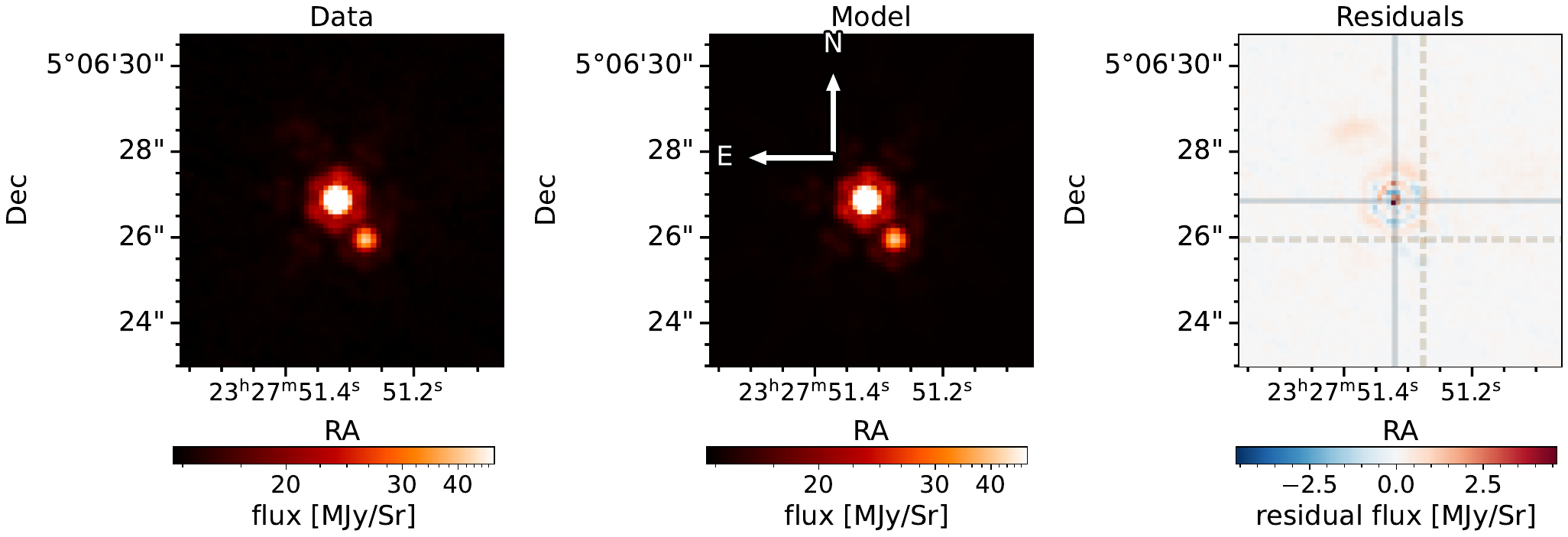}
    \caption{Two-source model of the candidate D F1000W image evaluated without the brighter-fatter correction (top row) and with it (bottom row). From left to right, each row shows the data, the model of Equation~\ref{eq:bfcorr}, and the residual in the same reference frame as presented in Figure~\ref{fig:MIRI_images}. The top row uses the bare \texttt{stpsf} point spread functions, while the bottom row uses the fitted Voigt broadening. Note the different colour scalings and the fact that the colour scale of the top row is compressed by a factor of two relative to its peak so that the structure remains visible. Faint blue and orange lines on the residual panels mark the column and row cuts plotted in Figure~\ref{fig:resids}.}
    \label{fig:bfc_compare}
\end{figure*}

\begin{figure}
    \centering
    \includegraphics[width=0.95\columnwidth]{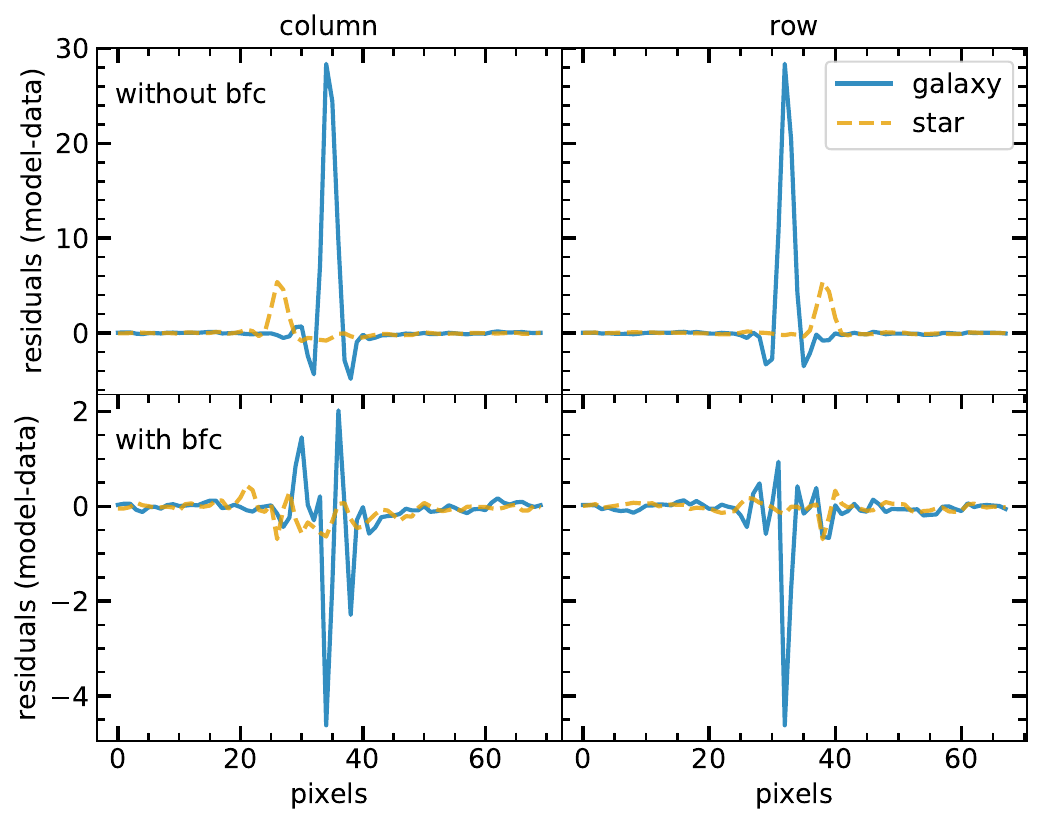}
    \caption{1D cuts through the residual images of Figure~\ref{fig:bfc_compare}. The left (right) column shows the cut taken along an image column (row) passing through the fitted centre of each source. The top row corresponds to the model without the brighter-fatter correction and the bottom row to the model with it. In each panel, the solid blue line is the cut through the background galaxy while the dashed orange line shows the cut through the M dwarf. The vertical axis is the model minus the data in MJy~sr$^{-1}$. The strong oscillation within a few pixels of each centroid in the top row is reduced by an order of magnitude in the bottom row.}
    \label{fig:resids}
\end{figure}

Figure~\ref{fig:bfc_compare} demonstrates the effect of the brighter-fatter correction defined in Section~\ref{sec:SED_modelling} on the candidate D F1000W fit. We take the posterior median of the eleven-parameter fit, evaluate the model of Equation~\ref{eq:bfcorr}, both with and without the fitted Voigt broadening, and display the data, the model, and the residual for each case. Without the correction, the residual carries a strong structured pattern in the form of a central deficit characteristic of an over-peaked model core failing to match the broadened data, with residuals reaching as high as $\sim28$~MJy~sr$^{-1}$. With our empirical brighter-fatter correction applied, the residuals fall to approximately $5$~MJy~sr$^{-1}$, consistent with the noise. The faint coloured lines on the residual panels indicate where the 1D cuts of Figure~\ref{fig:bfc_compare} are taken.

Figure~\ref{fig:resids} shows the same residuals as 1D cuts through the horizontal and vertical lines shown in Figure~\ref{fig:bfc_compare}. That is, using the fitted centroids, we sample the residual along the image row and column that pass through each source and plot the model minus the data against position. The left column of the figure corresponds to the column cut and the right column to the row cut; the top row shows the model without the brighter-fatter correction. Within every panel, the solid blue line follows the galaxy and the dashed orange line the star. The pronounced oscillation seen within a few pixels of each centroid in the top row is the 1D counterpart of the ringed residual in Figure~\ref{fig:bfc_compare}, and its disappearance in the bottom row confirms that the fitted broadening absorbs the brighter-fatter excess and removes the corresponding flux bias.

\begin{figure}
    \centering
    \includegraphics[width=0.99\columnwidth]{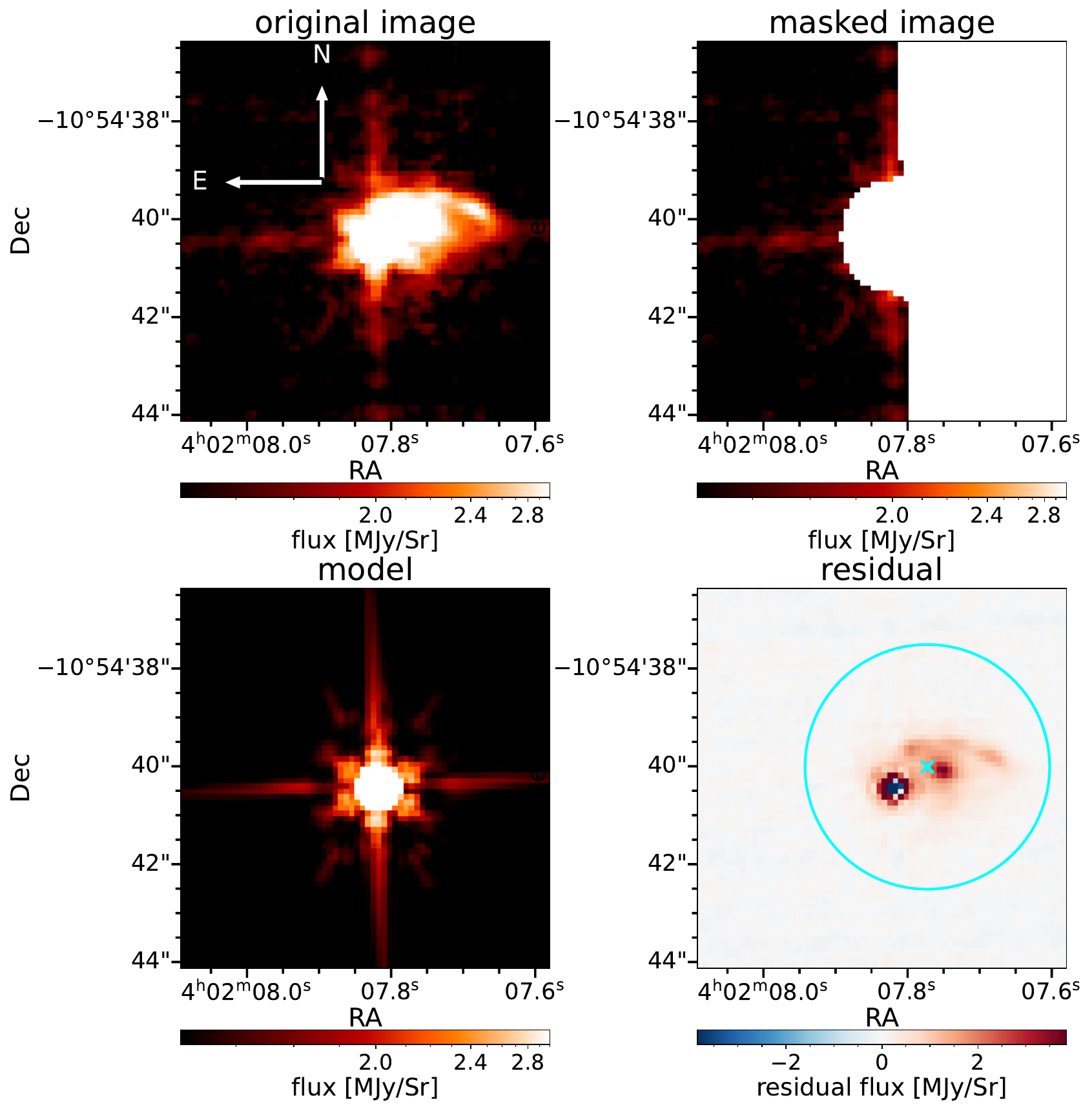}
    \caption{Recovery of the candidate E stellar flux in F560W from its diffraction spikes. The panels show, from left to right and top to bottom, the original image, the masked image used to put tight priors on flux and position, the best-fitting single-source model of the star, and the residuals after the star has been subtracted. The mask removes the half of the field that contains the galaxy together with a disk of radius $1.1''$ that is centered on the brightest pixel, leaving only the spikes that extend to the north, south, and east. The cyan circle and cross on the residual panel mark the aperture, centered on the galaxy, within which the residual galaxy flux is summed.}
    \label{fig:spike_phot}
\end{figure}

Figure~\ref{fig:spike_phot} illustrates how the F560W flux of candidate E is recovered from the diffraction spikes, as outlined in Section~\ref{sec:SED_modelling}. The four panels (from left to right and top to bottom) show the original image, the masked image that was used to set priors on the flux positions, the model of the star, and the residual that remains once the star is subtracted. We build the mask by discarding the half of the field that holds the galaxy along with a disk of radius $1.1''$ that is centered on the brightest pixel of the retained half. This removes the galaxy and the blended stellar core, leaving only the spikes reaching north, south, and east into comparatively empty sky. We use the diffraction spikes together with a series of least-squares fits to place tight priors on flux and position. That is, we allow the flux to vary between the limits where we just start to over- and under-subtract the diffraction spikes relative to the background. We use these priors to perform a single-source fit with \texttt{dynesty}, sampling the stellar centroid, flux, and background. We fix the Voigt broadening term to the same parameters fit for the M dwarf in field D at $5.6\mu$m, a choice that is justified by the fact that this star has a similar well-depth and groups per integration as candidate D (see, e.g., \citealt{Argyriou23}). The residual panel reveals the galaxy that is then measured within the aperture drawn on it to obtain the galaxy flux listed in Table~\ref{table:photometry}.

Table~\ref{table:bfc_fits} collects the recovered fluxes and centroids for the three fits that we consider reliable, namely the candidate D fits at F560W and F1000W and the candidate E fit at F560W, together with a fit to the candidate D F560W and F1000W images in which the brighter-fatter term is switched off. We list the AB magnitude of each component, obtained by summing the model surface brightness over the source and converting through the image zero point, as well as the centroid in ICRS coordinates. The error columns build up a realistic error budget, where we have added in quadrature the $1\sigma$ widths of our \texttt{dynesty} posteriors, a $2\%$ absolute flux calibration term \citep{Gordon2025}, correlated resampling noise, and a PSF mismatch term. Importantly, Table~\ref{table:bfc_fits} reveals that switching the brighter-fatter correction off leaves centroids unchanged at the sub-milliarcsecond level but moves the recovered star in F1000W from $18.53$ to $18.68$ and the galaxy from $16.57$ to $16.65$, which is the photometric counterpart of the residual structure removed in Figures~\ref{fig:bfc_compare} and \ref{fig:resids}.

\begin{table*}
\centering
\caption{Fluxes and centroids recovered by the \texttt{dynesty} fits for the three reliable cases, with the candidate D F560W and F1000W fits repeated without the brighter-fatter (BF) correction to expose its effect on the photometry. Magnitudes are on the AB system and centroids are in the ICRS. The magnitude error is broken into the formal posterior width $\sigma_{\rm stat}$, that width inflated for correlated resampling noise $\sigma_{\rm noise}=f_{\rm corr}\,\sigma_{\rm stat}$ with $f_{\rm corr}\approx2.6$ to $2.8$ measured from blank-sky apertures, and the point spread function term $\sigma_{\rm PSF}$ from refitting over a grid of \texttt{stpsf} models spanning jitter and wavefront date. The total $\sigma_{\rm tot}$ adds these in quadrature with an absolute flux calibration term of $2$ per cent or $0.022$ magnitudes \citep{Gordon2025}. The column $\sigma_{\rm pos}$ is the total centroid uncertainty, the formal posterior scatter inflated by $f_{\rm corr}$ and combined in quadrature with the centroid spread across the PSF grid. The final column is the reduced chi-square of the image fit, evaluated with the pipeline per-pixel errors used in the fit and listed once per fit. Values above unity are driven mainly by the correlated noise that those per-pixel errors do not capture, with only a minor contribution from residual PSF mismatch, and the brighter-fatter correction lowers them substantially. The candidate E galaxy is measured by aperture photometry rather than by the sampler and so is not listed here.}
\label{table:bfc_fits}
\footnotesize
\setlength{\tabcolsep}{4pt}
\begin{tabular}{llccccccccc}
\hline
Fit & Component & RA (deg) & Dec.\ (deg) & $\sigma_{\rm pos}$ (mas) & $m_{\rm AB}$ & $\sigma_{\rm stat}$ & $\sigma_{\rm noise}$ & $\sigma_{\rm PSF}$ & $\sigma_{\rm tot}$ & $\chi^2_\nu$ \\
\hline
D, F560W (with BF)   & star   & $351.9636514$ & $+5.1072079$  & $0.67$ & $17.509$ & $0.002$  & $0.005$ & $0.026$ & $0.034$ & $2.41$ \\
                     & galaxy & $351.9638365$ & $+5.1074600$  & $1.72$ & $18.735$ & $0.006$  & $0.016$ & $0.026$ & $0.038$ &        \\
D, F560W (no BF)     & star   & $351.9636514$ & $+5.1072079$  & $1.22$ & $17.608$ & $0.001$  & $0.003$ & $0.028$ & $0.036$ & $7.71$ \\
                     & galaxy & $351.9638361$ & $+5.1074599$  & $2.13$ & $19.489$ & $0.002$  & $0.007$ & $0.109$ & $0.111$ &        \\
D, F1000W (with BF)  & star   & $351.9636514$ & $+5.1072079$  & $1.04$ & $18.528$ & $0.006$  & $0.016$ & $0.038$ & $0.047$ & $1.91$ \\
                     & galaxy & $351.9638354$ & $+5.1074602$  & $0.61$ & $16.567$ & $0.001$  & $0.003$ & $0.019$ & $0.029$ &        \\
D, F1000W (no BF)    & star   & $351.9636514$ & $+5.1072079$  & $1.02$ & $18.675$ & $0.002$  & $0.004$ & $0.048$ & $0.053$ & $7.79$ \\
                     & galaxy & $351.9638352$ & $+5.1074600$  & $0.75$ & $16.646$ & $<0.001$ & $0.001$ & $0.034$ & $0.041$ &        \\
E, F560W (with BF)   & star   & $60.5325741$  & $-10.9112376$ & $1.71$ & $17.179$ & $0.001$  & $0.002$ & $0.021$ & $0.030$ & $12.84$ \\
\hline
\end{tabular}
\end{table*}

We build the error terms presented in Table~\ref{table:bfc_fits} as follows. The statistical width $\sigma_{\rm stat}$ is the $1\sigma$ width of the \texttt{dynesty} posterior. Because the mosaics are resampled, neighbouring pixels are correlated and the pipeline noise estimates are too low for any summed quantity, so we measure the true noise from the scatter of blank-sky apertures the size of the PSF and inflate $\sigma_{\rm stat}$ to $\sigma_{\rm noise}$ accordingly. The reduced chi-square in the final column instead uses the pipeline per-pixel errors directly and lies above unity, reflecting the same correlated noise on a per-pixel scale, where the errors are too low by only about its square root, with a small added contribution from residual PSF mismatch. We do not rescale this chi-square by $f_{\rm corr}$, which characterises integrated rather than per-pixel noise. The term $\sigma_{\rm PSF}$ captures how much the photometry depends on the assumed PSF. We regenerate the model PSF over a grid of plausible pointing jitters and of wavefront measurements taken around the time of each observation, refit every source with each one, and take the spread of the recovered magnitudes, which is $\sim0.02$ to $0.05$ mag. Finally, $\sigma_{\rm cal}$ is the absolute flux calibration of MIRI imaging, for which \citet{Gordon2025} find a repeatability near $1\%$ and we adopt a conservative $2\%$.

The formal precision of the \texttt{dynesty} fits, which lie at the millimagnitude and sub-milliarcsecond level, is far smaller than the realistic uncertainties that are dominated by the PSF model and the absolute flux calibration. The PSF term is the least certain of these, and, although its typical $1\sigma$ spread is $\sim0.02$ to $0.05$ mag, the most discrepant PSFs move the flux by as much as $\sim0.15$ mag, a shift that is comparable to the one seen when the brighter-fatter correction is removed. This means our fits are only as trustworthy as our \texttt{stpsf} model PSF, and a single choice of PSF can therefore bias the photometry at the $0.01$ to $0.15$ magnitude level, partially motivating the conservative $0.1$ magnitude error floor that we adopted in the main text. However, such examples are outliers, and $\sigma_{\rm PSF}$ thus resembles the inner $68\%$ spread incurred across our PSF grid. Still, our comparisons confirm that the correction removes a real bias rather than a statistical fluctuation, shifting the magnitudes by the amounts quoted above and lowering the F1000W reduced chi-square from $7.8$ to $1.9$, and, for the faint candidate D galaxy at F560W, it also tightens the photometry, cutting $\sigma_{\rm PSF}$ from $0.11$ to $0.03$ mag.


\bsp	
\label{lastpage}
\end{document}